\documentclass[aps,prl,twocolumn,showkeys]{revtex4-2}

\usepackage{graphicx}
\graphicspath{{Figures/}}

\usepackage{wrapfig}
\usepackage{float}
\usepackage[utf8x]{inputenc}
\usepackage{amsmath}
\usepackage{amssymb}
\usepackage{amsthm}
\usepackage{mathtools}
\usepackage{color}
\usepackage{multirow}
\usepackage{tabularx}
\usepackage{comment}
\usepackage{soul}
\usepackage{siunitx}
\DeclareSIUnit\bar{bar}

\begin{document}
\preprint{APS/123-QED}

\title{WISP Searches on a Fiber Interferometer under a Strong Magnetic Field}

\author{Josep Maria Batllori}
\email{Corresponding author: josep.batllori.berenguer@uni-hamburg.de}

\author{Yikun Gu}

\thanks{Present address: Centro de Astropartículas y Física de Altas Energías (CAPA), Universidad de
Zaragoza, 50009 Zaragoza, Spain}

\author{Dieter Horns}

\author{Marios Maroudas}

\author{Johannes Ulrichs}

\affiliation{Institut für Experimentalphysik, Universität Hamburg, Luruper Chaussee 149, D-22761 Hamburg, Germany}

\date{\today}

\begin{abstract}
A novel table-top experiment is introduced to detect photon-axion conversion: WISP Searches on a Fiber Interferometer (WISPFI). The setup consists of a Mach-Zehnder-type interferometer with a fiber placed inside an external magnetic field, where mixing occurs which is detected by measuring changes in amplitude. Hollow-core photonic crystal fibers (HC-PCF) will be used to achieve resonant mixing that is tuneable by regulating the gas pressure in the fiber.  An unexplored axion mass-range (\SIrange[range-phrase = --]{28}{100}{\meV}) can be probed reaching the two-photon coupling expected for the QCD axion. 
\end{abstract}


\maketitle

\section{Introduction}
\label{sec:introduction}

Axions are weakly-interacting pseudoscalar particles introduced to solve the strong CP problem in quantum chromodynamics (QCD)
which have been identified to be a candidate for  Cold Dark Matter (CDM) \cite{peccei_mathrmcp_1977, kim_weak-interaction_1979, dine_simple_1981, Sikivie_experimental_1983}. The QCD axion with mass $m_a$ inherits a non-vanishing two-photon coupling-strength
$g_{a\gamma\gamma}\propto m_a$ that is model dependent. This two-photon coupling leads to a rich phenomenology that can be explored both experimentally as well as observationally. While cosmological and astrophysical searches are sensitive to a wide range of the axion parameter space, laboratory experiments searching for axions as CDM (so-called haloscopes) have achieved so far the best sensitivity and start to rule-out the benchmark QCD axion models for a narrow mass-range from \SIrange{2.81}{3.31}{\micro\eV} \cite{admx_collaboration_extended_2020}. However, these results depend upon the local density of CDM, which is poorly constrained and could be substantially smaller than the average at similar galactocentric distances \cite{eggemeier_minivoids_2022}.
On the other hand, laboratory experiments that do not rely on axions to form CDM (e.g. light-shining-through-wall \cite{Sikivie_experimental_1983}), or searches for birefringence \cite{Cameron_optic_1993} are less sensitive and none of the existing (and projected) experiments achieve sufficient sensitivity to probe the QCD axion (for an overview see e.g. \cite{zyla_particle_2020}). 

\begin{figure*}[!htb]
    \centering
    \includegraphics[width=\linewidth]{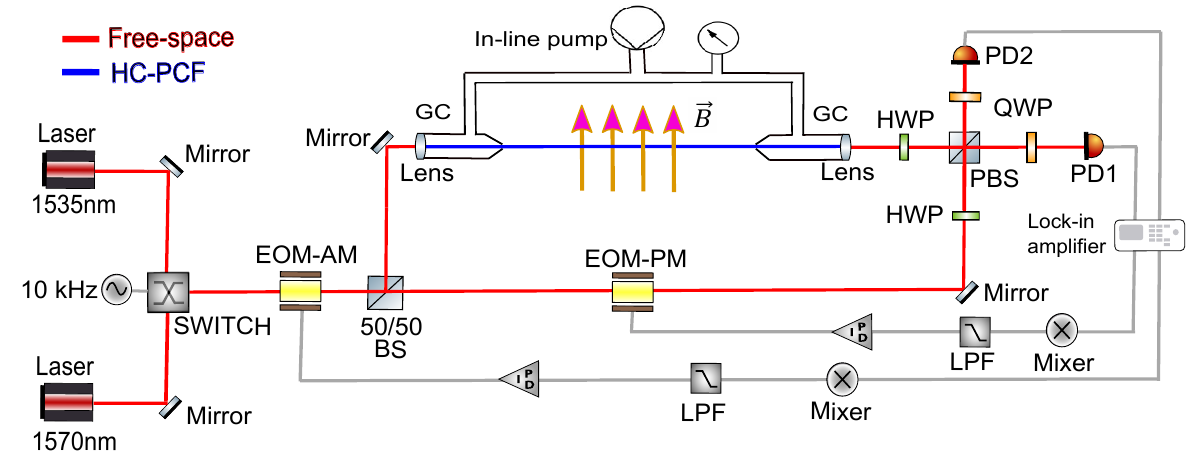}
    \caption{Schematic view of the experimental setup of WISPFI considering a partial-free space MZI for detecting photon-axion oscillations. In red, the laser beam in free space is shown, while in blue the light beam propagating through the HC-PCF is represented. The latter is the sensitive arm of the interferometer and is placed inside a \SI{9}{\tesla} dipole magnetic field with a length of \SI{100}{\meter} where the photon-axion resonant mixing occurs. The various acronyms correspond to: electro-optical modulator (EOM), beam-splitter (BS), gas cell (GC), half-waveplate (HWP), polarized beam-splitter (PBS), quarter-waveplate (QWP), photo-detector (PD), and low-pass filter (LPF). More details about the setup are given in the main text.}
    \label{fig:general_setup}
\end{figure*}

In this paper, we introduce a new experimental setup called WISPFI (WISP searches on a Fiber Interferometer) that focuses on photon-axion conversion in a waveguide by measuring photon reduction in the presence of a strong external magnetic field \cite{tam_production_2012}. In this novel approach, light guiding over long distances can be achieved together with resonant detection inside the bore of a strong magnet. The basic idea of WISPFI is to use a Mach-Zehnder type interferometer (MZI) where a laser beam is split into two arms with one arm used as a reference and the other arm placed inside a strong magnetic field which induces a photon-to-axion conversion (see Fig.~\ref{fig:general_setup}, further details are given in a later Section). Then, an amplitude reduction can be measured in the presence of a non-vanishing photon-axion coupling $g_{a\gamma\gamma}$. The measurable effect of axion-photon mixing relies on the  Primakoff effect. 
The resulting conversion probability \cite{raffelt_mixing_1988} $P_{ \gamma \rightarrow a}\propto g_{a\gamma\gamma}^{2}(BL)^{2}\ll 1$, where $g_{a\gamma\gamma}$ is the axion-photon coupling coefficient, $BL$ is the product of the transversal magnetic field $B$ and the length $L$ that the photon beam passes through the external magnetic field. As a comparison, in light-shining-through-wall experiments, the signal rate depends on the product of photon-to-axion and axion-to-photon conversion probabilities which therefore scales $P_{\gamma \rightarrow a \rightarrow \gamma}\propto g_{a\gamma\gamma}^{4}(BL)^{4}\ll P_{\gamma\rightarrow a}$.

\section{Photon-axion mixing in  hollow-core photonic crystal fibers (HC-PCF)}
\label{sec:resonant_mixing_hcf}

The photon-to-axion conversion probability assuming a mode propagating in the z-direction \cite{raffelt_mixing_1988} is:
\begin{equation}
\label{eq:pconv4}
    P_{\gamma \rightarrow a}=\sin^{2}(2\theta) \sin^{2}\left(k_\mathrm{osc} z\right),
\end{equation} 
where $k_\mathrm{osc}$ is the oscillation wave-number (see Appendix for more details on the photon-axion mixing). The first term in Eq.~\ref{eq:pconv4} accounts for the amplitude of the oscillations while the second term accounts for the oscillations. The mixing angle $\theta$ is deduced from the diagonalisation of the mixing matrix 
with the off-diagonal term $G=g_{a\gamma\gamma} B/2$ in Eq.~\ref{eq:diagonalise} in the Appendix:
\begin{equation}
\label{eq:mixing_angle}
  \tan(2\theta)=\frac{G}{\Delta_{-}},
\end{equation} 
where $\Delta_{-}=(k_{\gamma}^{2}- k_{a}^{2})/{4\omega}$ is the photon-axion transfer momenta, given in terms of the photon with energy $\omega$, $k_{\gamma}$, and axion wave-momenta, $k_{a}$, propagating in the core-media of the fiber. In the following, we will focus on the
conversion at or close to the resonant condition such that $\Delta_-\approx 0$, where the resulting probability $P_{\gamma\rightarrow a}$ is energy-independent and the oscillation wave-number simplifies to $k_\mathrm{osc}=G$. Under the assumption of $P_{\gamma\rightarrow a}\ll 1$, Eq.~\ref{eq:pconv4} can be simplified and evaluated for typical values for the magnetic field and length of the fiber:
\begin{equation}
\label{eq:presonant}
    P_{\gamma \rightarrow a} = 8\times10^{-19}
    \left(\frac{g_{a\gamma\gamma}}{10^{-12}~\si{\GeV\tothe{-1}}}\right)^2 
    \left(\frac{B}{\SI{9}{\tesla}}\right)^2
    \left(\frac{z}{\SI{100}{\meter}}\right)^2.
\end{equation}

\subsection{Resonant conversion}

\begin{figure}[!htb]
    \centering
    \includegraphics[width=0.4\linewidth]{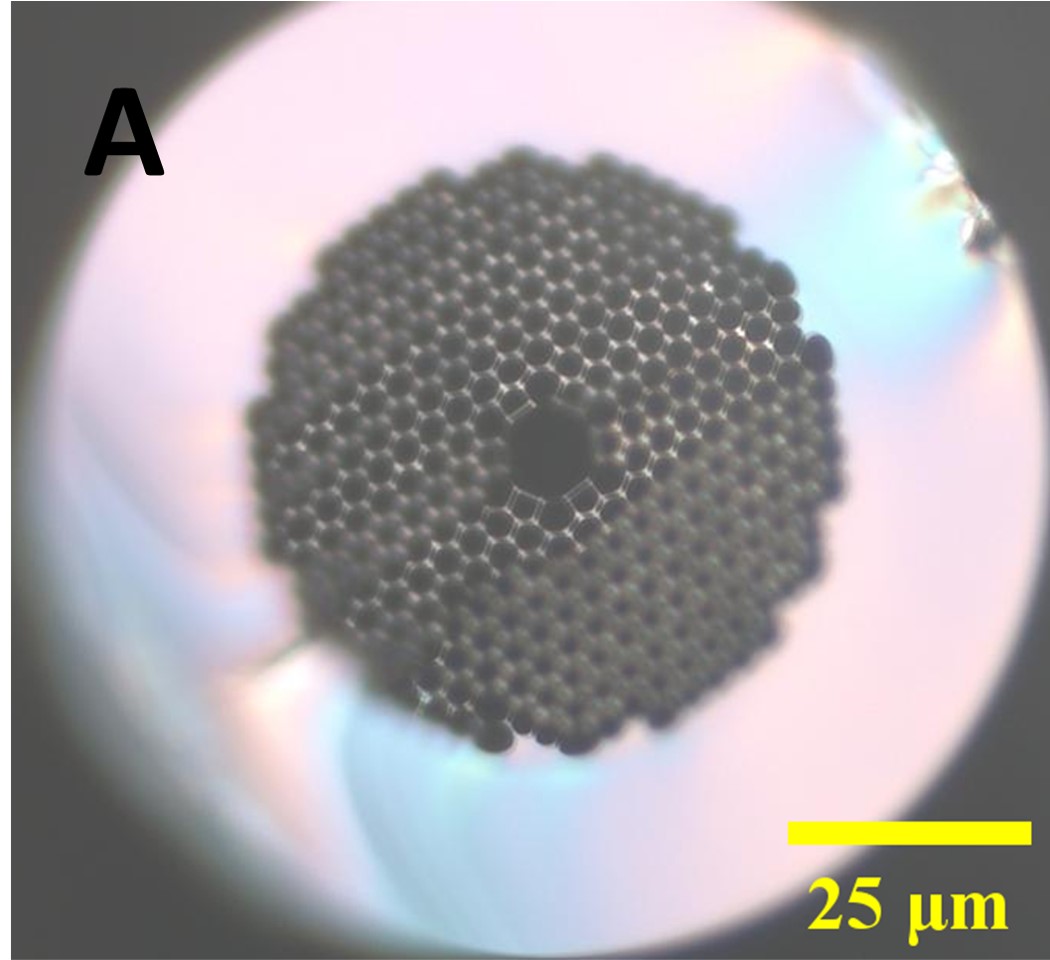}
    \includegraphics[width=0.52\linewidth]{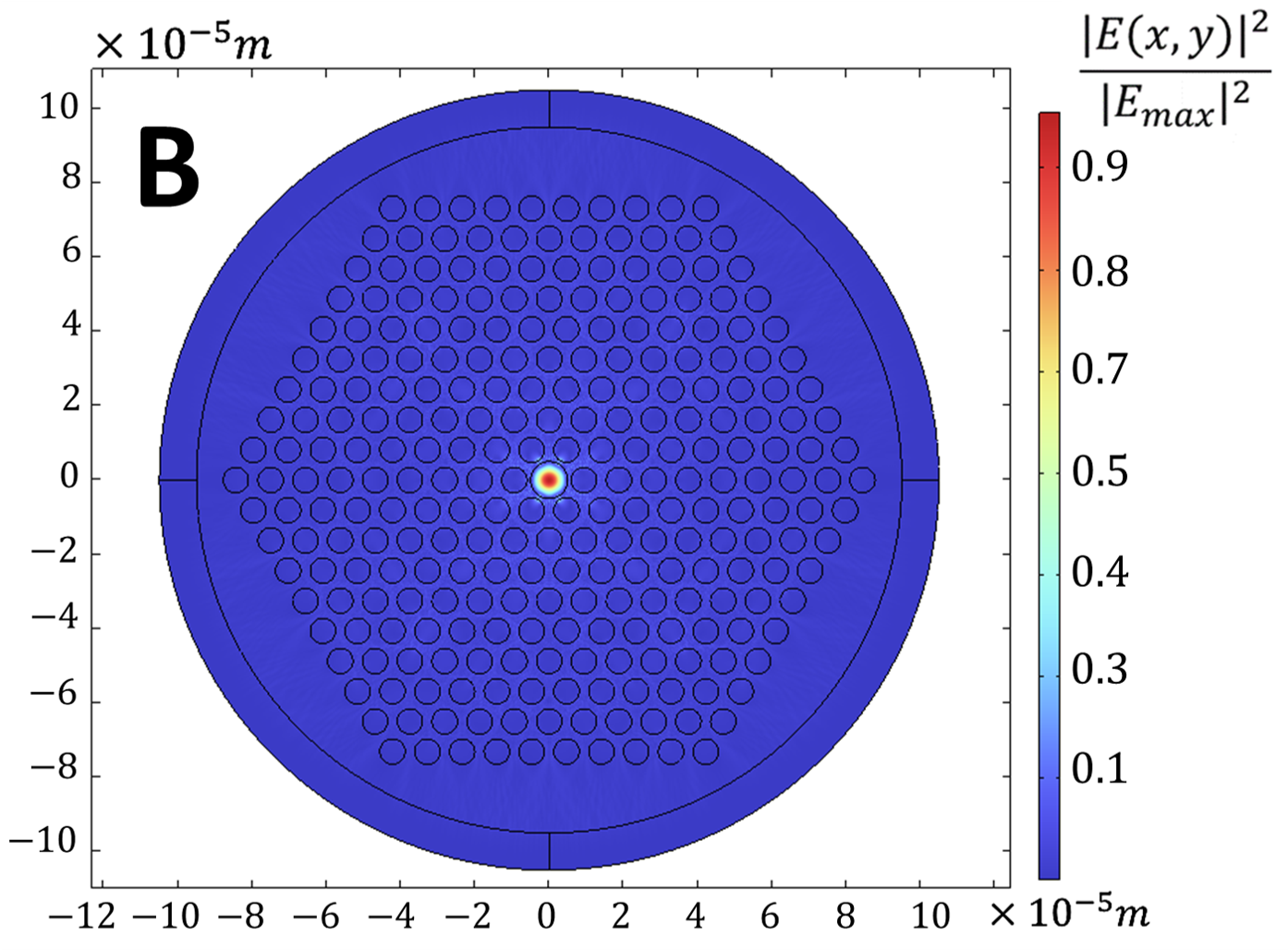}
    \caption{\textbf{A}: Image taken with a OLYMPUS MX40 microscope of a commercial HC-PCF (HC-1550) with \SI{5}{\micro\meter} core radius. \textbf{B}: FEM simulation of a mode field distribution in a HC-PCF with \SI{5}{\micro\meter} core radius, a capillary-to-core radius ratio of 0.682 and a wavelength of \SI{1.55}{\micro\meter}. The calculated effective mode index is 0.992 for a pressure of \SI{0.1}{\bar}.}
    \label{fig:hcf}
\end{figure}

The resonant condition $\Delta_{-} = 0$ is asymptotically achieved for large energy $\omega$ or equal momenta for the photon and axion: $k_{\gamma}=k_{a}$. At resonance, the mixing angle from Eq.~\ref{eq:mixing_angle} is $45^{\circ}$ which maximizes the amplitude term in Eq.~\ref{eq:pconv4}. The resonant conversion occurs in a medium with effective refractive index $n_\mathrm{eff}$ for an axion mass $m_a$ given by:
\begin{equation}
\label{eq:axion_mass}
    m_{a}=\omega\sqrt{1-n_\mathrm{eff}^{2}}.
\end{equation}
This condition can not be fulfilled for waveguides based on dielectric materials. However, the resonance condition can be fulfilled by using a HC-PCF \cite{cregan_single_1999, russel_photonic_2003}, which is a particular type of optical fiber with exceptional applications in detection and sensing \cite{nikodem_laser_2020}. The light is guided through a low-refractive index hollow core, which can be filled either with a gas or a fluid. The core is surrounded by a hexagonal periodic arrangement of holes in the cladding, generating the photonic bandgap structure of the material. See Fig.~\ref{fig:hcf}A for a microscopic view of the cross-section of a commercially available HC-PCF. Additionally, HC-PCFs have a higher damage threshold for the guided laser power compared to standard step-index fibers \cite{Laurent_2004}. By exploiting the bandgap structure of HC-PCF, the propagating mode can acquire a refractive index below 1 (see Eq.~\ref{eq:neff_formula} in the Appendix) which based on Eq.~\ref{eq:axion_mass} leads to real axion masses at resonant mixing.

\subsection{Effective mode index of HC-PCF}
\label{sec:hcf_analytical}

From Eq.~\ref{eq:axion_mass} we can observe that the resonance occurs for an axion mass which changes with the refractive index of the propagating mode. Here, we investigate the effective mode index for a HC-PCF configuration (see Fig.~\ref{fig:hcf}B) by solving the Maxwell equations with the finite-element method (FEM) with COMSOL.

The effective mode index of the ground mode in a HC-PCF depends on parameters such as the core radius $R_c$, pressure $p$, and wavelength $\lambda$. In turn, the resonant conversion is limited to an axion mass close to the value given in Eq.~\ref{eq:axion_mass}. 
We consider the difference $\Delta n:=1-n_\mathrm{eff}$ and consequently for 
\begin{equation}
\label{eq:axion_mass_n_eff}
    m_a = \omega\sqrt{(1-n_\mathrm{eff})(1+n_\mathrm{eff})} \approx \omega \sqrt{2\Delta n},
\end{equation}
for $\Delta n \ll 1$.

\begin{figure}[!htb]
    \centering
    \includegraphics[width=\linewidth]{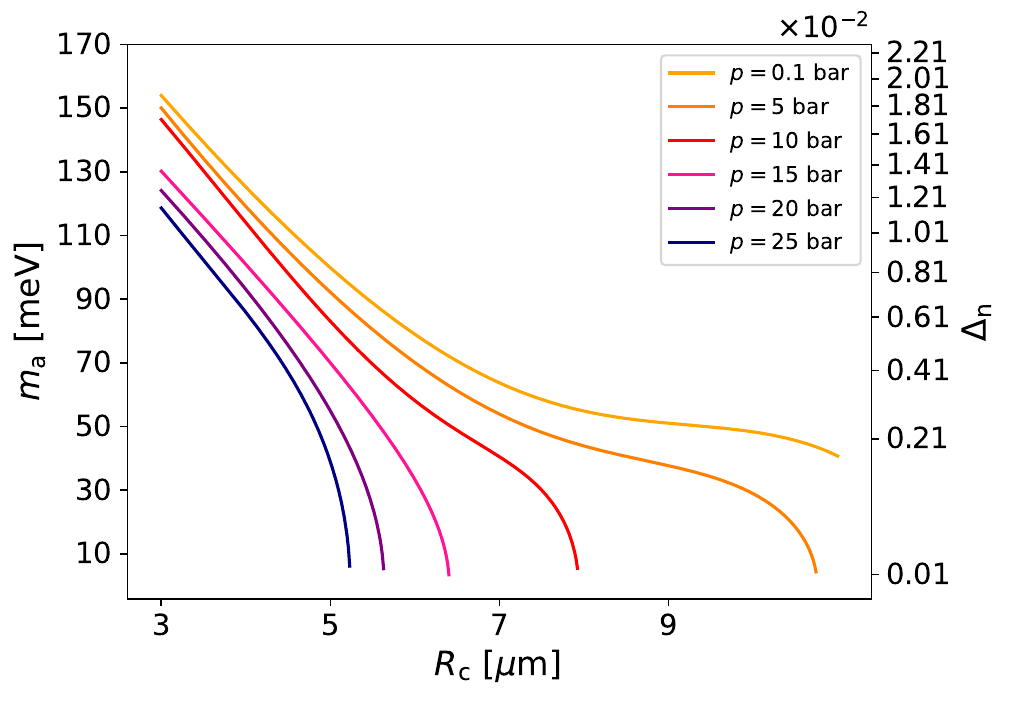}
    \caption{Axion mass at resonance and the difference of the effective mode index ($\Delta n=1-n_\mathrm{eff}$) as a function of the core radius of the HC-PCF for different pressures of air filling the hollow core, using FEM solver. An average wavelength of \SI{1.55}{\micro\meter} is applied. The assumed temperature of the air inside the core of the fiber is \SI{20}{\celsius}.}
    \label{fig:effective_mode_index}
\end{figure}

The results from FEM simulations for a varying core radius $R_c$ and pressure $p$ are shown in Fig.~\ref{fig:effective_mode_index}. The probed axion masses for resonant conversion based on different core radii and different pressures of air that fill the hollow core of the fiber vary between $m_a \approx \SIrange{10}{160}{\meV}$. The observed increase of the effective mode index with increasing core radius or pressure matches the analytical approximation (see Eq.~\ref{eq:neff_formula} in the Appendix). These simulations were performed assuming a straight fiber.

\section{Overview of the experimental setup}
\label{sec:setup_overview}

The general setup for the WISPFI experiment is based on a partial free space MZI with the sensing arm implemented by a waveguide that can be easily integrated inside a magnetic bore. Two lasers at different wavelengths of \SI{1535}{\nm} and \SI{1570}{\nm} are used together with an optical switch changing between the two lasers to modulate the axion signal at a desired frequency ($\approx \SI{10}{\kHz}$). This way, a modulation amplitude close to $100\%$ can be achieved (see also Eq.~\ref{eq:sensitivity_baseline_setup}). The wavelengths of the two lasers are chosen to be sufficiently separated from each other to guarantee that at most one laser is converting axions in resonance while the other one is necessarily off-resonance. Therefore, the possible observed axion-generated signal will be modulated with the selected frequency of the optical switch. This approach is similar to the one suggested by \cite{tam_production_2012} in order to isolate the photon-axion conversion signal at side-bands with frequencies $\omega_m$, $2\omega_m$. Next, an electric optical modulator (EOM-AM) is used to compensate for the expected variations in amplitude caused by the optical switch without a magnetic field present. A free space BS is then used for splitting equally the beam and redirecting it to the two interferometer arms accordingly. The axion-photon oscillation leads to a relative amplitude loss and phase shift of the sensing arm embedded in the magnetic field with respect to the reference arm of the interferometer. It is noted, that for resonant conversion the axion-induced phase shift is zero \cite{raffelt_mixing_1988}. 

In the setup shown in Fig.~\ref{fig:general_setup}, the sensing arm is made of a HC-PCF and is placed in a strong magnetic field of $\SI{9}{\tesla}$ where the photon-axion conversion takes place in resonant conditions. The HC-PCF can be optionally pressurized to change the mass range for resonant conversion as demonstrated in Fig.~\ref{fig:effective_mode_index}. The control of the conditions of the medium inside of the HC-PCF is one of the major advantages of this type of waveguide and has been demonstrated in several experimental setups \cite{cao_fiber_2014, triches_portable_2015}. 

The reference arm of the Mach Zehnder interferometer is realized in free space with an electro-optical modulator (EOM-PM). The working point of the interferometer is then locked using a PID-driven EOM-PM which compensates for any possible phase changes between the two arms up to frequencies of a few \si{\MHz}. At the end of both arms, a PBS is implemented in free space. This way, we achieve the necessary control to match and balance the interferometer arms. For proper control over the full $2\pi$ polarization and manipulation of the beam an adjustable HWP and QWP are placed before and after the PBS, respectively, as shown in Fig.~\ref{fig:general_setup}. 

The side-bands are then detected using PD1 operating in the dark fringe. The signal is further demodulated following a lock-in amplifier scheme as shown in Fig.~\ref{fig:general_setup} with a mixer and a LPF. The set point is carefully chosen by observing the resulting fringe pattern from performing a sweep of the offset voltage in amplitude. In addition, the slope of a particular fringe provides the sensitivity of the interferometer to changes in the optical path length. As previously mentioned the interferometer is then stabilized by computing the error signal as the difference between the measured amplitude and the set point and sending it back to the EOM-PM. In addition, the signal from the bright port of the MZI where PD2 is placed is sent back to the EOM-AM via a PID loop (see Fig.~\ref{fig:general_setup}).

\section{Sensitivity analysis}
\label{sec:sensitivity}

The expected sensitivity of the WISPFI setup as introduced in the preceding Section is estimated under the following assumptions: (i) The MZI is operated at a dark fringe, (ii) the instrumental noise is dominated by the dark current of the PD, (iii) there are no additional losses included. The signal-to-noise ratio (SNR) is then given by the ratio of the power received in the dark port for a conversion probability in resonance (see Eq.~\ref{eq:presonant}) and the noise induced by the dark current. For an uncooled InGaAs photodiode, the noise-equivalent power (NEP) amounts to approximately \SI{0.5}{\femto\watt\per\sqrt{\Hz}}. At the dark fringe, the influence of shot noise coming from the signal in terms of NEP can be estimated as $\mathrm{NEP_{SN}}=\sqrt{P_{tot}\cdot P_{\gamma \rightarrow a}\cdot\omega}$. For an axion - photon coupling of $g_{a\gamma\gamma}=1.2\times10^{-12}~\si{\GeV^{-1}}$ and $P_\mathrm{tot}=\SI{4}{\watt}$, the shot noise contribution from the signal is expected to be $\mathrm{NEP_{SN}}\approx 6\times10^{-4}~\si{\femto\watt\per\sqrt{\Hz}}$, which is significantly smaller than the noise contribution from the PD. However, for coupling strength values $g_{a\gamma\gamma}\ge 1\times10^{-9}~\si{\GeV^{-1}}$, the shot noise dominates over the detector-noise and thus limits the resulting sensitivity. An additional noise contribution term is derived from the shot noise from the laser integrated over the sideband. Assuming a bandwidth of $\approx \SI{100}{\hertz}$ for the sideband and a residual power of less than $1\%$ received in the dark fringe, the resulting noise contribution is smaller than the detector noise.

For a commercial laser with $P_\mathrm{tot}=\SI{4}{\watt}$ and $\lambda = \SI{1550}{\nm}$, a fiber length of $L=\SI{100}{\meter}$ embedded into a \SI{9}{\tesla} dipole magnetic field of equal length, and an operation time of \SI{180}{\day}, the resulting sensitivity on $g_{a\gamma\gamma}$ is given by:
\begin{equation}
\label{eq:sensitivity_baseline_setup}
\begin{aligned}
    g_{a\gamma\gamma} &\approx 1.2\times10^{-12}\si{\GeV\tothe{-1}} \left(\frac{\mathrm{SNR}}{3}\right)^{1/2}
    \left(\frac{B}{\SI{9}{\tesla}}\right)^{-1} \\
   &\left(\frac{L}{\SI{100}{\meter}}\right)^{-1}
    \left(\frac{P_\mathrm{tot}}{\SI{4}{\watt}}\right)^{-1/2}
    \left(\frac{\beta_\mathrm{sig}}{1}\right)^{-1/2} \\
   &\left(\frac{t}{\SI{180}{\day}}\right)^{-1/4}
    \left(\frac{\mathrm{NEP_{SN+PD}}}{\SI{0.5}{\femto\watt\per\sqrt{\Hz}}}\right)^{1/2}.
\end{aligned}
\end{equation}
It is noted, that for the calculation of the sensitivity, the fiber is assumed to be straight and the polarization is maintained along the whole path so that the E-field is parallel to the B-field to maximize the photon-axion conversion. This can be achieved with Polarization-Maintaining (PM) HC-PCF \cite{fini_polarization_2014, chen_single_2016}. A very low polarization cross-coupling can be achieved without significant birefringence effects using anti-resonant fibers (ARFs) \cite{Taranta_polarization_purity_2020}. Additionally, the effective mode index is calculated for a constant temperature of \SI{20}{\celsius} and a constant pressure of \SI{0.1}{\bar} along the \SI{100}{\meter}-long HC-PCF. Under these conditions, the effective mode index remains constant along the fiber.

The mass for resonant conversion in the HC-PCF (see Eq.~\ref{eq:axion_mass}) is mainly determined by the core radius $R_c$ (see Eq.~\ref{eq:neff_formula}). The production process of a HC-PCF leads to random variations of the core radius along the fiber. Since the fiber is drawn from a heated mandrel, initial variations are stretched such that the largest variations of core radii occur on the longest length scale of the resulting fiber. The resulting variations resemble a $1/f$ or red noise behaviour with a cut-off at a length scale of \SI{10}{\meter}.

\begin{figure}[!htb]
    \centering
    \includegraphics[width=\linewidth]{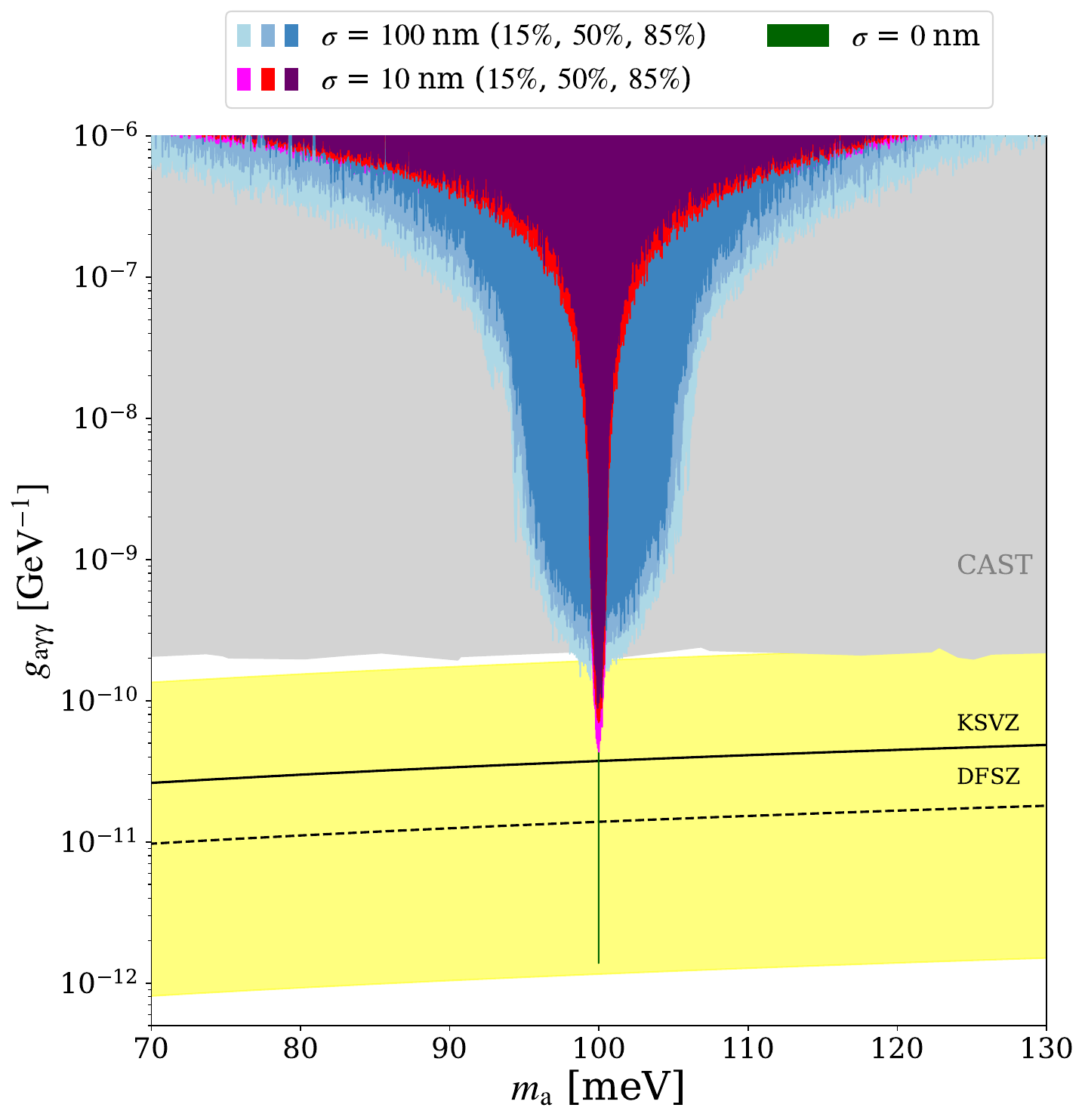}
    \caption{Projected sensitivity for WISPFI experiment with an external magnetic field of \SI{9}{\tesla}. In this baseline setup, the applied laser power is \SI{4}{\watt} in a HC-PCF fiber with a length of \SI{100}{\meter}, and a core radius with a mean value of \SI{5}{\micro\meter} and $\sigma = \SI{100}{\nano\meter}$, \SI{10}{\nano\meter}, and \SI{0}{\nano\meter}. An average laser wavelength of \SI{1.55}{\micro\meter} is applied. The total data-taking time is assumed to be 6 months for each value of $\sigma$. The 15, 50, and 85 percentiles are also shown in different colors. The CAST limit is also shown for comparison \cite{CAST_new_2017}.}
    \label{fig:exclusion_plot_baseline}
\end{figure}

In order to characterize the impact of the random variations introduced at the time of drawing the fiber, we assume a mean value for $R_c$ of \SI{5}{\micro\metre} with $\sigma = \SI{100}{\nano\meter}$, \SI{10}{\nano\meter}, and \SI{0}{\nano\meter}. The resulting sensitivity is then determined from a simulation of 10 fiber realizations each with a \SI{100}{\meter} length. In each case, the resulting conversion probability $P_{\gamma\rightarrow a}$ is calculated using the transfer-matrix approach applied to 1000 segments each with \SI{0.1}{\meter} length (see \cite{mirizzi_stochastic_2009, deangelis_relevance_2011}). This way, we can calculate for each realization the range of resonant mass and coupling that can be probed. In Fig~\ref{fig:exclusion_plot_baseline}, the 15, 50, and 85 percentile are shown in different colors. With the projected sensitivity, for $\sigma = \SI{10}{\nano\meter}$ and 50 percentile, the WISPFI experiment will probe the $g_{a\gamma\gamma}$ coupling for the QCD axion close to the KSVZ scenario within 6 months. The probed axion mass range can be varied between $m_a\approx\SIrange{10}{100}{\meV}$ by changing the gas pressure inside the core of the fiber between \SIrange{0.1}{30}{\bar} (see also Fig.~\ref{fig:pressure}). In the case of a wider variation of core radii in the fiber considered in Fig.~\ref{fig:exclusion_plot_baseline}, the axion mass range will extend without additional tuning, however, with decreased sensitivity.

As discussed in a previous Section, the probed axion mass at resonance can be tuned by changing the air pressure inside the HC-PCF. The tuning step size is determined by the Full Width at Half Maximum (FWHM) of the resonance and therefore, in our setup, it will be adjusted based on the actual core radius variations of the used HC-PCF which will also determine the reached sensitivity. More details on the pressure variations along the HC-PCF and the gas filling time are given in the appendix.

\section{Discussion}

\begin{figure}[!htb]
    \centering
    \includegraphics[width=\linewidth]{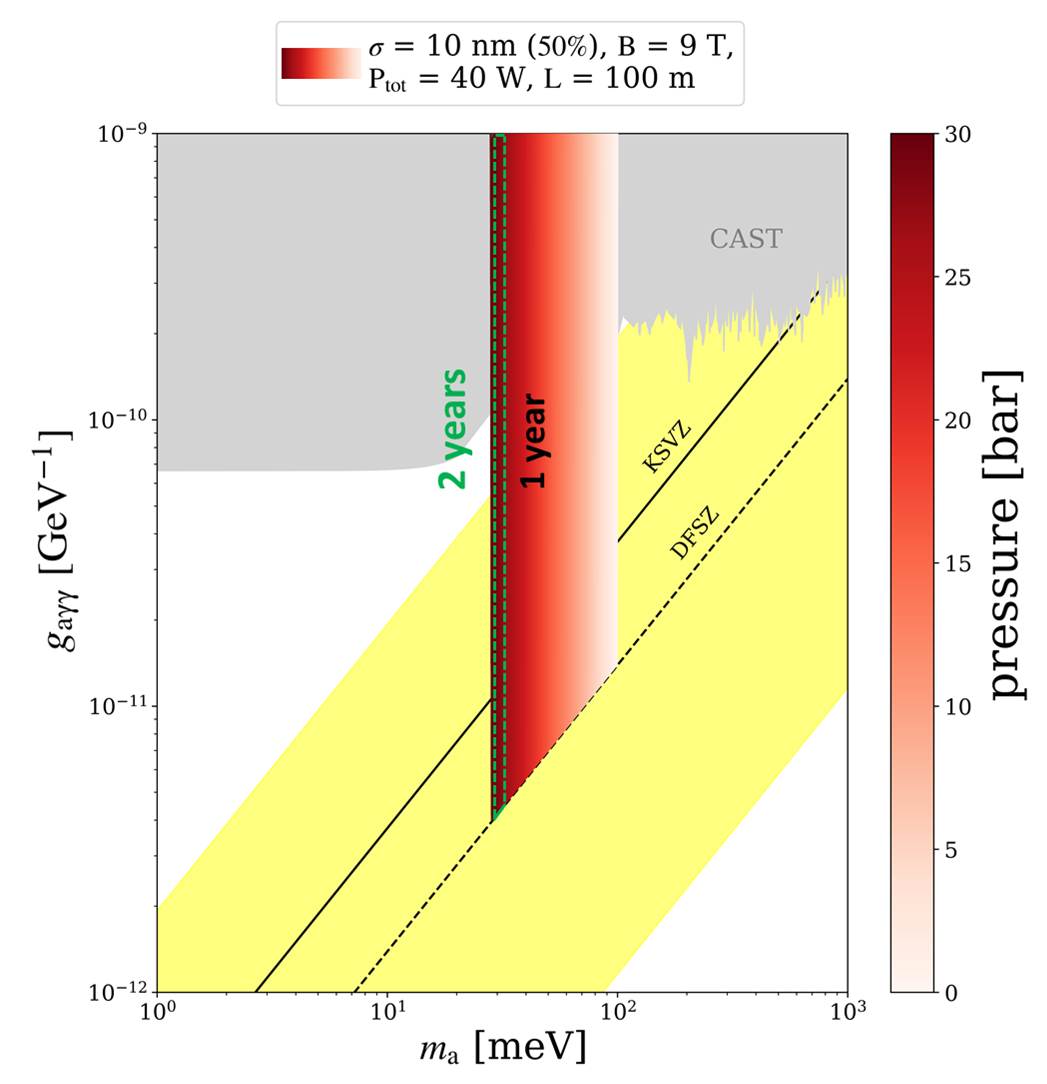}
    \caption{WISPFI's long-term prospects including a \SI{40}{\watt} laser, a \SI{100}{\m}-long HC-PCF fiber with a core radius of \SI{5}{\micro\meter} and $\sigma=\SI{10}{\nm}$, and a Fabry-Pérot cavity with a finesse of $F=100$. An average laser wavelength of \SI{1.55}{\micro\meter} is applied. The data-taking time is 1 year and 2 years (green-dotted region). The 50 percentile chosen for this case, results in DFSZ sensitivity. The tuning is performed by controlling the air pressure in the core of the HC-PCF from \SIrange{0.1}{27.9}{\bar}. This results in an axion mass range between $\sim \SIrange{28}{100}{\meV}$. The CAST limit is also shown for comparison \cite{CAST_new_2017}.}
    \label{fig:exclusion_plot_longterm}
\end{figure}

The WISPFI experiment will be sensitive to the QCD axion in a narrow mass range close to \SI{100}{\milli\eV} that is so far unexplored experimentally. The baseline configuration presented above (Fig.~\ref{fig:exclusion_plot_baseline})
can be substantially improved by installing several optimized interferometers tuned individually to a resonant conversion for different axion masses (e.g., by choosing
gas pressure, wavelength, and core radii). The resulting sensitivity and mass range covered is therefore scaleable without the need to develop new methods.  
The novel approach to search for resonant conversion in a HC-PCF opens additional unique opportunities: By attaching electrode strips to the fiber, it is possible to probe the photon-axion conversion in strong electric fields that has been predicted in scenarios based upon modified Quantum-Electromagnetodynamics \cite{sokolov_electromagnetic_2022}. Additionally, a Fabry-Pérot cavity can be used to increase the effective power on the sensitive arm and thus improve the resulting constraint on $g_{a\gamma\gamma}$ by $F^{1/2}$, where $F$ is basically the finesse of the cavity \cite{tam_production_2012}. Such hollow core fiber Fabry-Pérot interferometers have been shown to exhibit a finesse of over 3000 \cite{Ding_fabry_2020}. For the locking of the cavity, the Pound-Drever-Hall technique will be used. It is noted that the modulation frequency of the optical switch is carefully selected such that the decay time of the cavity is shorter than the switching time between the two laser wavelengths. An example, using a laser of \SI{40}{\watt}, a \SI{100}{\m}-long HC-PCF with $\sigma=\SI{10}{\nm}$, and a Fabry-Pérot cavity with a moderate finesse of 100, while measuring for a total of 1 and 2 years is shown in Fig.~\ref{fig:exclusion_plot_longterm}. For the case of 2 years, a total pressure change of \SI{27.8}{\bar} is applied in $210$ steps of approximately \SI{132}{\milli\bar} corresponding to $\sim \SI{0.6}{\meV}$ in mass, in order to reach axion masses between $\sim \SIrange{28}{100}{\meV}$ with a DFSZ sensitivity.

\section{Summary and outlook}

We have introduced a MZI embedded in an external magnetic field that is sensitive to the conversion of photons to axions. The conversion takes place resonantly in a HC-PCF. By changing the gas pressure in the hollow core, the refractive index for the guided mode can be tuned. In the approach presented, the resulting sensitivity reaches the QCD axion band for the photon-axion coupling $g_{a\gamma\gamma}$ in a so-far unexplored mass range at \SI{100}{\milli\eV}. The approach is scalable and we have introduced several avenues to improve the sensitivity and mass range further.

\section*{Acknowledgements}

This project is funded by the Deutsche Forschungsgemeinschaft (DFG, German Research Foundation) under Germany’s Excellence Strategy – EXC 2121 ''Quantum Universe" – 390833306, and through the DFG funds for major instrumentation grant DFG INST 152/824-1. YG acknowledges funding from the Chinese Scholarship Council Scholarship No. 201708060377. We thank Oliver Gerberding, Rebecca Harte, Le Hoang Nguyen, and Martin Tluczykont for their useful comments, discussions, and contributions to the WISPFI prototype development. We also thank Joerg Jaeckel, Giuseppe Ruoso, Andreas Ringwald, and Georg Raffelt for pointing out the difference in the photon and axion propagation in a coiled fiber. We acknowledge initial discussions with Roman Schnabel and Pascal Gewecke on using squeezed light for WISPFI. We appreciate discussions and support from Zurich Instruments on the application of UHFLI on our prototype. Finally, we gratefully acknowledge Michael Frosz and Philip Russell for their valuable discussions on HC-PCFs and the fiber sample contribution to our prototype development. This article is based upon work from COST Action COSMIC WISPers CA21106, supported by COST (European Cooperation in Science and Technology).

\section{Appendices}

\subsection{Equation of motion}

From the axion-modified Ampere and Faraday laws, the Axion-Maxwell-Helmholtz equation focusing on the z-direction of propagation can be deduced as:
\begin{equation}
\label{eq:maxwell-axion-coupled}
   \left[\partial^{2}_{z}+M^2\right]\psi(z)=0,
\end{equation} 
where $\psi(z)=\begin{pmatrix} E(z) \\ a(z) \end{pmatrix}$ being the electric field and the axion field, respectively. The variables $\omega_{\gamma}$ and $\omega_{a}$ are the angular frequencies of the photon and axion fields, respectively, which will be assumed to be the same in the following calculations. In the ultra-relativistic limit $|m|^{2}\ll \omega^2$, Eq.~\ref{eq:maxwell-axion-coupled} is linearized in the context of observing plane-waves, as done in Ref~\cite{raffelt_mixing_1988}. Regarding that, the mixing matrix is
\begin{equation}
\label{eq:mixing_matrix}
M^2=\frac{1}{2\omega}\begin{pmatrix} k^2_{\gamma} & -G \\ -G & k^2_{a} \end{pmatrix},
\end{equation} 
where $k_{\gamma}^{2}=n_{eff}^{2}k_{o}^{2}=\omega_{\gamma}^{2}-m_{\gamma}^{2}$ and $k_{a}^{2}=\omega_{a}^{2}-m_{a}^{2}$ give the photon and axion momenta respectively in the core media of the fiber, and $G=g_{a\gamma\gamma}B/2$ is the mixing energy which is proportional to the external magnetic field and the coupling strength. To obtain the proper propagation constants of the photon and axion fields a change of basis is required to decouple the system. For that reason, we work in a new basis $(\tilde{E}(z),\tilde{a}(z))$ so that both fields can be treated independently and the standard Maxwell boundary conditions of continuity and differentiability can be applied.  By finding the mixing angle $\theta$ that decouples the fields and makes the mixing matrix diagonal, the eigenvalues $\tilde{k}_{\gamma, a}$ can be estimated. We then have all the information for dealing with each field separately.

\begin{equation}
\label{eq:diagonalise}
  \tilde{M}^2=\begin{pmatrix} \tilde{k}_{\gamma}^{2} & 0 \\ 0 & \tilde{k}_{a}^{2}   \end{pmatrix}=U(\theta)\frac{M^2}{2\omega}U^{-1}(\theta),
\end{equation} 
where $\tilde{k}_{\gamma, a}=\Delta_{+}\pm\sqrt{\Delta_{-}^{2}+G^{2}}$ with $\Delta_{\pm}=(k_{\gamma}^{2}\pm k_{a}^{2})/{4\omega}$. The mixing angle $\theta$ can be obtained from the non-diagonal terms of the resulting matrix operation in Eq.~\ref{eq:diagonalise}:

\begin{equation}
\label{eq:mixing_angle_append}
  \tan(2\theta)=\frac{G}{\Delta_{-}}
\end{equation}

Note that the mixing angle in Eq.~\ref{eq:mixing_angle_append} has a dependence on the refractive index of the media. 

\subsection{Photon-Axion conversion probability in a fiber}

The photon-axion conversion probability inside a fiber will now be estimated. In the simple linearized case, the resulting decoupled electric and axion fields propagating inside the fiber in the z-direction can be expressed as $\tilde{E}(z)\sim \exp{(i\tilde{k}_{\gamma}z)}$ and $\tilde{a}(z)\sim \exp{(i\tilde{k}_{a}z)}$, respectively. The original fields can be then expressed in terms of the rotated and their respective factorized amplitude coefficients ($A_{\gamma}, A_{a}$):

\begin{equation}
\label{eq:basis}
    {\psi}(z)=\begin{pmatrix} E(z) \\ a(z) \end{pmatrix}=\begin{pmatrix} A_{\gamma} \cdot \tilde{E}(z) \\ A_{a} \cdot \tilde{a}(z) \end{pmatrix}.
\end{equation}
Given the total energy of the system and losing one degree of freedom by normalizing the amplitude coefficients, the total intensity can be expressed as $A_{\gamma}^{2}+A_{a}^{2}=I_{T}$, being $A_{i}$ real amplitude coefficients. We can then observe how our system conserves the total amplitude. The electric and axion fields in the $z$ direction can be therefore deduced by applying the rotation back to our original basis.

\begin{equation}
\label{eq:basisback}
\small
\begin{split}
    &{\psi}(z)=\begin{pmatrix} E(z) \\ a(z) \end{pmatrix}=U^{-1}(\theta)\tilde\psi(z)U(\theta)
\end{split}
\end{equation} 

Regarding that, the resulting conversion probability is:

\begin{equation}
\label{eq:pconv3}
    P_{\gamma \rightarrow a}=\frac{A_{a}^2}{I_{T}}=\cos^{2}\theta \sin^{2}\theta |e^{-i\tilde{k}_{\gamma}z}-e^{-i\tilde{k}_{a}z}|^{2}.
\end{equation}

This expression can be further simplified by describing it in terms of the oscillation wavenumber $k_{osc} = \tilde{k}_{\gamma} - \tilde{k}_{a}$ as
\begin{equation*}
    P_{\gamma \rightarrow a}=\sin^{2}(2\theta) \sin^{2}\left(k_\mathrm{osc} z\right),
\end{equation*}
which corresponds to Eq.~\ref{eq:pconv4} in the main text.

\subsection{Estimation of the effective mode index}
\label{appx:estimation_neff}

To analytically calculate the effective mode index in a HC-PCF the fiber-core is commonly approximated considering a circular capillary, expressing the mode profile as in the case of a hollow waveguide \cite{marcatili_hollow_1964}. Under that approach, only the core mode is relevant for the conversion. To ensure good confinement and simplification of the calculation, we assume that the ratio between the core radius and the wavelength in vacuum is much higher than one ($R_{c}/\lambda\gg 1$).

The effective mode index can then be calculated as in Eq.~\ref{eq:neff_formula} from the real part of the propagation constant, $k_{\gamma}$, divided by the photon wave-number in vacuum, $k_{o}$. We only consider the real (loss-less) part of the propagation constant since optical fibers are normally made of dielectric materials with a predominantly real refractive index. The effective mode index ($n_\mathrm{eff}$), and eventually the axion mass, are principally affected by the core radius ($R_{c}$) or by changing the refractive index of the filled gas ($n_{gas}$), by varying pressure ($p$), temperature ($T$), and wavelength ($\lambda$) \cite{marcatili_hollow_1964}: 
\begin{equation}
\label{eq:neff_formula}
    n_\mathrm{eff}=\frac{k_{\gamma}}{k_{o}}=\sqrt{n^{2}_\mathrm{gas}(\lambda, p, T)-\left(\frac{u_\mathrm{nm}}{k_{o}R_{c}}\right)^{2}},
\end{equation}
where $u_\mathrm{nm}$ is the $m^{th}$ zero of the $n^{th}$-order Bessel function of the first kind. This so-called Marcatili's formula is generally valid when the ratio between the core radius and the applied wavelength is $R_{c}/\lambda\gtrsim 27$ \cite{Rosa_2021}. However, for smaller values of $R_c/\lambda$, it overestimates the resulting effective mode index. Because of that, we consider a more 
suitable model based on a hollow core surrounded by a ring of dielectric tubes. Each tube is separated from each other by a trapezoidal-shape gap as given in Ref.~\cite{Rosa_2021}. In this approximation, the core radius $R_c$ in Eq.~\ref{eq:neff_formula} is replaced by an effective core radius $R_\mathrm{eff}(\lambda)$ (see Eq.~10 of \cite{Rosa_2021}). This approximation is known as tube-lattice fiber (TLF). 

Since the actual geometry of the HC-PCF under study contains additional rings of hollow tubes, the effective mode index is also calculated through simulations where the cylindrical two-dimensional geometry of the system has been implemented in the commercial software package COMSOL-Multiphysics \cite{comsol_multiphysics} (see the black outline in Fig.~\ref{fig:hcf}B). Based on the available HC-PCF in the market (HC-1550, Fig.~\ref{fig:hcf}A) the hollow core radius is assumed to be $R_c = 5~\mu\mathrm{m}$ and the cladding  is composed of pure silica glass 
($n_\mathrm{clad} = 1.45$). We consider a capillary-to-core radius ratio of $0.682$~\cite{uebel_broadband_2016}. The cylindrical geometry is matched by a  boundary condition that corresponds to an infinite radial extension of the cladding. The resulting effective mode index is found by scanning the solution of the Maxwell equations for propagating modes that are confined to the core. As an example, for such a solution with the pressure and temperature fixed at $p=\SI{0.1}{\bar}$ and $T=\SI{20}{\celsius}$ respectively, we show in Fig.~\ref{fig:hcf}B in color-scale the electric field strength along the propagation direction which shows a clear maximum collocated with the hollow core.

\begin{figure}[!htb]
    \centering
    \includegraphics[width=\linewidth]{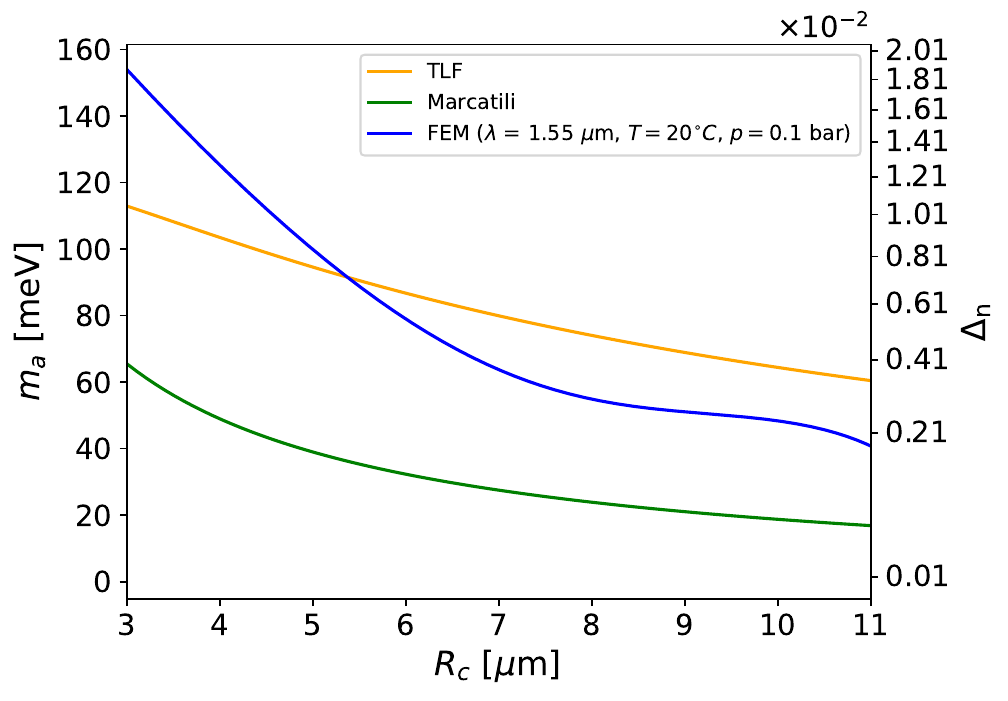}
    \caption{Axion mass at resonance and the difference of the effective mode index ($\Delta n=1-n_\mathrm{eff}$) as a function of the core radius for a pressure of \SI{0.1}{\bar} and a wavelength of \SI{1.55}{\micro\meter} using FEM solver (blue). A comparison is also shown with the TLF model \cite{Rosa_2021} (orange) and the Marcatili model \cite{marcatili_hollow_1964} (green). The corresponding error bars from the simulation which correspond to the numerical uncertainty on $n_\mathrm{eff}$ are on the level of $10^{-4}$ and are therefore not visible.}
    \label{fig:core}
\end{figure}

The solutions obtained with the FEM calculations with the core radius $R_c$ varying between \SIrange{3}{11}{\micro\meter} are shown in Fig.~\ref{fig:core} in comparison with the analytical solutions introduced above (see Eqs.~\ref{eq:neff_formula}, \ref{eq:axion_mass_n_eff}). The TLF approximation is closer to the numerical solution for core radii $\sim$\SI{5}{\micro\meter} and larger while the differences are more pronounced at smaller core radii as expected. The Marcatili-based solution shows a qualitatively similar behaviour but the overall value for $\Delta n$ is smaller. These differences with the FEM simulation are already discussed in \cite{finger_accuracy_2014}. The resulting range of $m_a$ varies between \SIrange{40}{160}{\milli\eV} for a pressure of \SI{0.1}{\bar}. It is noted that a smaller core radius leads to a smaller effective mode index (larger axion mass), as the mode experiences stronger interaction with the cladding.

\begin{figure}[!htb]
    \centering
    \includegraphics[width=\linewidth]{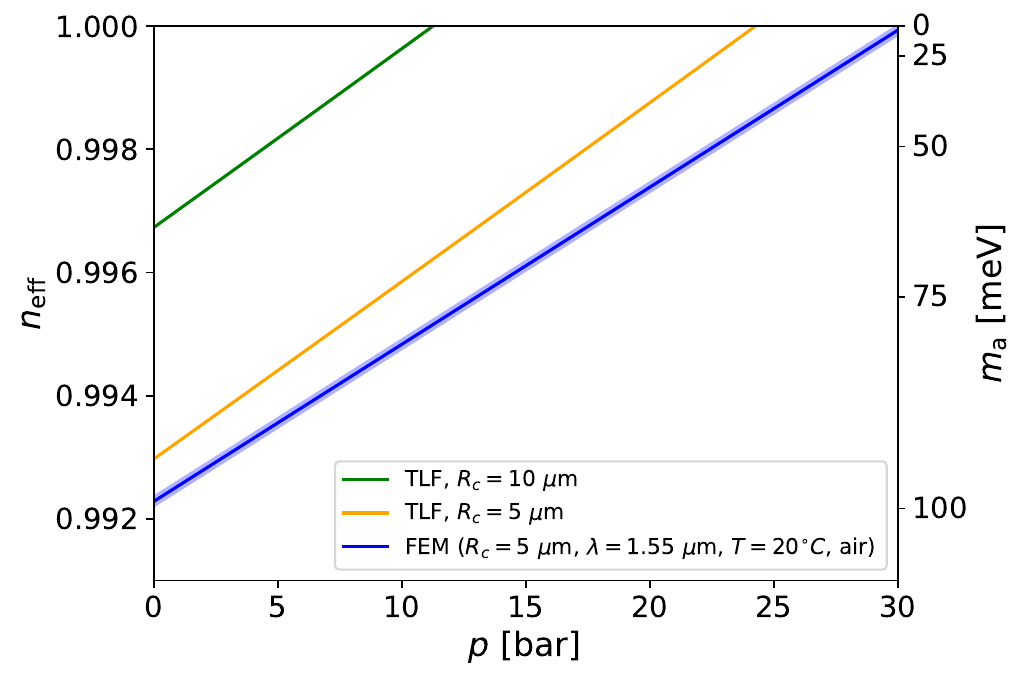}
    \caption{Effective mode index and axion mass at resonance as a function of pressure using FEM solver (blue). The core radius of the HC-PCF is assumed to be \SI{5}{\micro\meter}, while the applied laser wavelength is \SI{1.55}{\micro\meter} and the temperature of the air inside the core of the fiber is \SI{20}{\celsius}. The blue-shaded region corresponds to the numerical uncertainty of $10^{-4}$ in the effective mode index. The TLF model \cite{Rosa_2021} for a core radius of \SI{5}{\micro\meter} (orange) and \SI{10}{\micro\meter} (green) are also shown for comparison.}
    \label{fig:pressure}
\end{figure}

Besides changes in the geometry, the effective mode index is also affected by changes in the refractive index of the medium that fills the hollow core through changes in pressure and temperature. First, the refractive index for humid air is calculated using least squares to fit the raw data from \cite{Mathar_2007} under the standard conditions ($T_0=\SI{20}{\celsius}$, $p_{0}=\SI{0.1}{\bar}$, relative humidity $H_{0}=50\%$) for $\lambda=\SI{1.55}{\micro\meter}$. Then, as shown in Fig.~\ref{fig:pressure}, the resulting variation of the effective mode index $n_\mathrm{eff}$ can be simulated when varying the pressure between $p=\SI{0.1}{\bar}$ and $p=\SI{30}{\bar}$. As expected, $n_\mathrm{eff}\propto p$ and it reaches vacuum-like conditions ($n_\mathrm{eff} = 1$) for $p\approx \SI{30.3}{\bar}$ (FEM simulated, in blue). It is noted, that for a larger core radius of $R_c=\SI{10}{\micro\meter}$,
vacuum-like conditions are reached for $p\approx\SI{11}{\bar}$ (TLF, in green).

\begin{figure}[!htb]
    \centering
    \includegraphics[width=\linewidth]{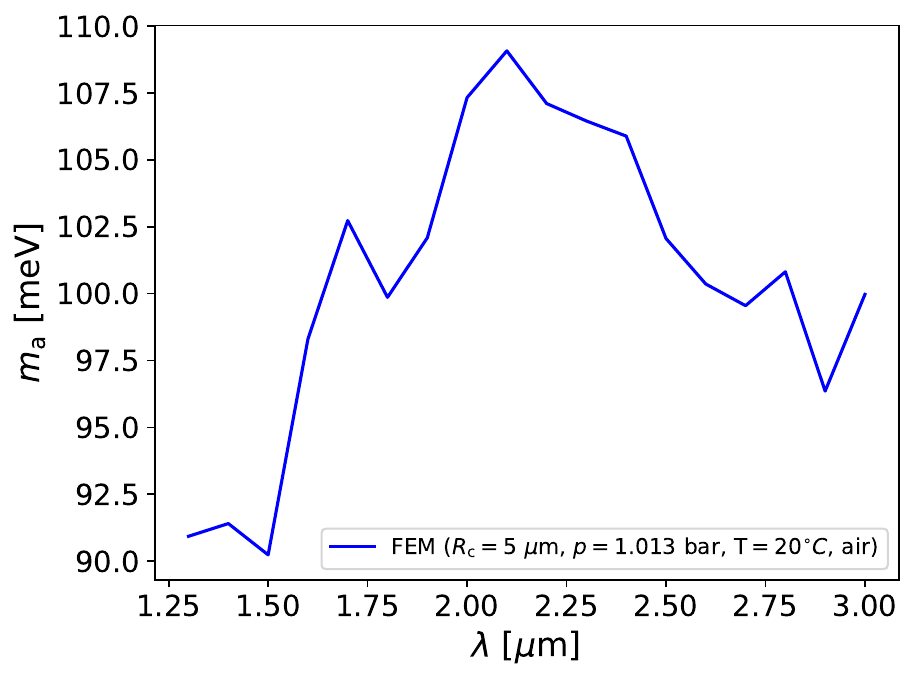}
    \caption{Axion mass as a function of the wavelength using FEM solver. The core radius of the HC-PCF is assumed to be \SI{5}{\micro\meter}, while the air inside the core of the fiber is at standard conditions.}
    \label{fig:wavelength}
\end{figure}

Finally, in Fig.~\ref{fig:wavelength}, the wavelength of the propagating light is varied to highlight the effect in the effective mode index of the HC-PCF and the subsequent probed axion mass. It should be mentioned, that longer wavelengths experience less confinement and a lower effective mode index, which can lead to a reduction in the overall transmission efficiency of the fiber. Furthermore, at larger wavelengths, the confinement loss can limit the transmission band of the core of the fiber. Analogously to the application of pressure changes, a fine-tuning of the wavelength can result in lower axion masses.

\subsection{Interferometric Detection}

The detection principle for photon-axion conversion in fiber is similar to the case of a Michelson interferometer setup as introduced by \cite{tam_production_2012}. The case of the amplitude-modulated laser beam is shown here as an example. For a laser beam passing through the EOM before entering the interferometer, a sinusoidal amplitude-modulated field can be expressed as:
\begin{equation}
\begin{aligned}
\label{eq:Ein1}
  E_\mathrm{in}
  &=E_{0}\left[
  1+\beta_\mathrm{sig}\sin(\omega_{m}t)
  \right]
  e^{i\omega t}\\
  &=E_{0}\left[
  e^{i\omega t} 
  +\frac{\beta_\mathrm{sig}}{2i} e^{i(\omega+\omega_{m})t} 
  -\frac{\beta_\mathrm{sig}}{2i} e^{i(\omega-\omega_{m})t} 
  \right]
\end{aligned}
\end{equation}
where $\beta_\mathrm{sig}$ is a constant for the relative modulation amplitude, and $\omega_{m}$ is the modulation frequency. In the second line of Eq.~\ref{eq:Ein1}, the first term indicates the carrier frequency and the latter two the sideband frequencies.

Let $L_\mathrm{sen}$ be the length of the sensing arm and $L_\mathrm{ref}$ the length of the reference arm, $L=(L_\mathrm{sen}+L_\mathrm{ref})/2$, $\Delta L=L_\mathrm{sen}-L_\mathrm{ref}$. The sideband signal can be maximized when manually choosing a specific macro-length difference between the two arms for the MZI of $k_{m}\Delta L/2 = \pi / 2$. 

By introducing  the photon-axion conversion $P_{\gamma \rightarrow a}$ in one of the arms of the interferometer, the output field can be deduced:
\begin{align}
\label{eq:Etot_a}
  E_\mathrm{tot} 
  &=\frac{E_{0}}{2} e^{i(\omega t+k_0L)}
  \Big[ 2i\sin(k_0\Delta L /2)
  -P_{\gamma \rightarrow a} e^{-ik_0\Delta L /2}  \nonumber \\
  &+\beta_\mathrm{sig}\big( 2\cos(k_0\Delta L /2)- P_{\gamma \rightarrow a} e^{-ik_0\Delta L /2}\big)\cdot \nonumber\\
  &\cdot\cos{(\omega_{m}t+k_{m}L)} \Big],
\end{align}
where $k_0$ and $\mathrm{k_m}$ are the wavenumbers for the laser beam and the modulated beam. 

It is noticed that the output power $|E_\mathrm{tot}|^2$ would have a DC component as well as higher order modulation components with frequency $\omega_{m}$ and $2\omega_{m}$. After mixing with a local oscillator, the component of frequency $\omega_{m}$ can be extracted through a low-pass filter. This leads to the $\omega_m$ modulated term (we only consider power loss and neglect the phase shift which is higher order in the coupling and disappears for the 
resonant conversion) from the total output power:
\begin{equation}
 \label{eq:Mod_Pterm}
  |E_m|^2 = -|E_0|^2\beta_\mathrm{sig}P_{\gamma \rightarrow a}\cos{(k_0\Delta L)}
   \cos{(\omega_{m}t+k_{m}L)}.
\end{equation}

This is the output signal of the sidebands from photon-axion mixing, where $|E_0|^2$ refers to the input laser power. When $k_0\Delta{L}=\pi /2$, dark fringe is achieved, and the power from the DC part is minimized.

\subsection{Fiber Pressurizing}

\begin{figure}[!htb]
    \centering
    \includegraphics[width=\linewidth]{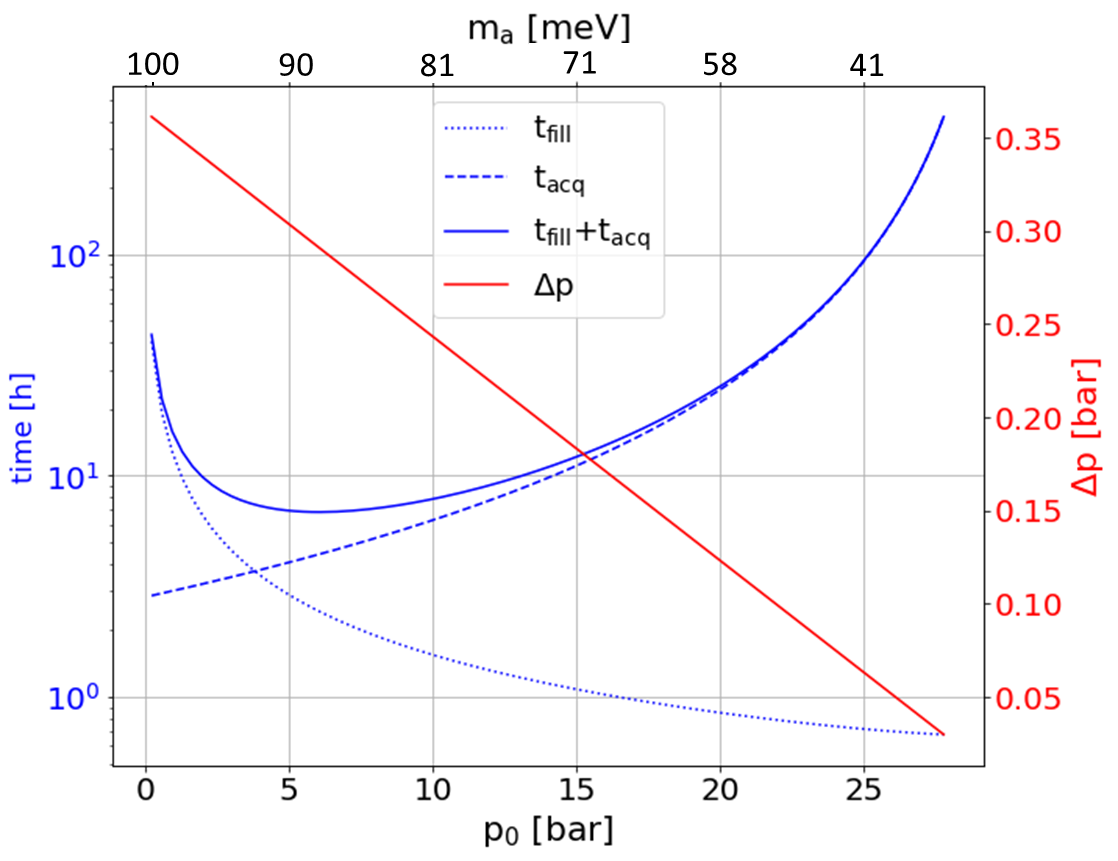}
    \caption{A comparison of the interval that is needed to achieve a pressure balance (less than $\Delta p$ variation of pressure along the fiber) 
    with the data acquisition time ($t_\mathrm{acq}$) required to reach the DFSZ sensitivity for different values of pressure $p_0$ (the corresponding mass range is indicated on the axis at the top). During data-taking, the pressure $p_0$ will be varied in the range from \SIrange{0.1}{27.8}{\bar}. The accumulated time to fill the fiber will be less than 2~\% of the total data acquisition time.}
    \label{fig:pressure_step}
\end{figure}

As mentioned previously, varying the pressure inside the core of the HC-PCF is the most efficient way for tuning the effective mode index and thus the probed axion mass. Pressuring of a HC-PCF using two gas cells at the two ends has already been demonstrated in the literature for gas sensing \cite{cao_fiber_2014, Masum_fiber_2019}. For a \SI{100}{\meter} fiber length the pressure is varied between \SIrange{0.1}{27.8}{\bar} to sample an axion mass between $\sim \SIrange{28}{100}{\meV}$. In Fig.~\ref{fig:pressure_step} the required pressure difference per tuning step is shown as it changes when increasing the pressure. The maximum pressure difference of $\approx \SI{0.35}{\bar}$ per tuning step is reached for the high-mass axion regime whereas it decreases to $\approx \SI{0.05}{\bar}$ for the low-mass axion regime. 

The required time to pressurize the HC-PCF depends on the gas that is used (dry air at T=$\SI{22}{\celsius}$), the length of the fiber, the required pressure, and the maximum pressure deviation at the midpoint of the fiber. We estimate the required filling time for a HC-PCF with a core radius of \SI{5}{\micro\meter} and length \SI{100}{\meter} following the approach suggested in \cite{Dicaire_pressurization_2010} that extrapolates between continuous and slip flow regime. The resulting filling time to achieve a pressure difference of less than $\Delta p$ at the midpoint of the fiber for a given external pressure $p_0$ is calculated using:

\begin{equation}
\label{eq:fill_time}
t_\mathrm{fill} = \frac{(L/2)^{2}}{\pi^2 D} \ln \left(\frac{p_0}{\Delta p}\frac{8}{\pi^2}\right),
\end{equation}
where $L/2$ is the midpoint of the HC-PCF fiber. 

For the assumed core radius variations of $\sigma = \SI{10}{\nano\meter}$, the FWHM of the projected sensitivity is $\sim \SI{0.6}{\meV}$, which translates into a limit on the pressure difference $\Delta p$ along the fiber of \SIrange{0.05}{0.35}{\bar} depending upon the pressure level. The resulting filling time ($t_\mathrm{fill}$) is also shown in Fig.~\ref{fig:pressure_step} as a function of pressure (\SIrange{0.1}{27.8}{\bar}) along with the specific time per step to reach the DFSZ theoretical line from Fig.~\ref{fig:exclusion_plot_longterm}.

It is noted that the total time for pressurizing the fiber over all 210 tuning steps is on the order of 13 days which is less than 2~\% of the required data-taking time to reach DFSZ sensitivity for the \SI{72}{\meV} axion mass range.

\bibliography{refs}

\begin{thebibliography}{32}%
\makeatletter
\providecommand \@ifxundefined [1]{%
 \@ifx{#1\undefined}
}%
\providecommand \@ifnum [1]{%
 \ifnum #1\expandafter \@firstoftwo
 \else \expandafter \@secondoftwo
 \fi
}%
\providecommand \@ifx [1]{%
 \ifx #1\expandafter \@firstoftwo
 \else \expandafter \@secondoftwo
 \fi
}%
\providecommand \natexlab [1]{#1}%
\providecommand \enquote  [1]{``#1''}%
\providecommand \bibnamefont  [1]{#1}%
\providecommand \bibfnamefont [1]{#1}%
\providecommand \citenamefont [1]{#1}%
\providecommand \href@noop [0]{\@secondoftwo}%
\providecommand \href [0]{\begingroup \@sanitize@url \@href}%
\providecommand \@href[1]{\@@startlink{#1}\@@href}%
\providecommand \@@href[1]{\endgroup#1\@@endlink}%
\providecommand \@sanitize@url [0]{\catcode `\\12\catcode `\$12\catcode
  `\&12\catcode `\#12\catcode `\^12\catcode `\_12\catcode `\%12\relax}%
\providecommand \@@startlink[1]{}%
\providecommand \@@endlink[0]{}%
\providecommand \url  [0]{\begingroup\@sanitize@url \@url }%
\providecommand \@url [1]{\endgroup\@href {#1}{\urlprefix }}%
\providecommand \urlprefix  [0]{URL }%
\providecommand \Eprint [0]{\href }%
\providecommand \doibase [0]{https://doi.org/}%
\providecommand \selectlanguage [0]{\@gobble}%
\providecommand \bibinfo  [0]{\@secondoftwo}%
\providecommand \bibfield  [0]{\@secondoftwo}%
\providecommand \translation [1]{[#1]}%
\providecommand \BibitemOpen [0]{}%
\providecommand \bibitemStop [0]{}%
\providecommand \bibitemNoStop [0]{.\EOS\space}%
\providecommand \EOS [0]{\spacefactor3000\relax}%
\providecommand \BibitemShut  [1]{\csname bibitem#1\endcsname}%
\let\auto@bib@innerbib\@empty
\bibitem [{\citenamefont {Peccei}\ and\ \citenamefont
  {Quinn}(1977)}]{peccei_mathrmcp_1977}%
  \BibitemOpen
  \bibfield  {author} {\bibinfo {author} {\bibfnamefont {R.~D.}\ \bibnamefont
  {Peccei}}\ and\ \bibinfo {author} {\bibfnamefont {H.~R.}\ \bibnamefont
  {Quinn}},\ }\href {https://doi.org/10.1103/PhysRevLett.38.1440} {\bibfield
  {journal} {\bibinfo  {journal} {Physical Review Letters}\ }\textbf {\bibinfo
  {volume} {38}},\ \bibinfo {pages} {1440} (\bibinfo {year}
  {1977})}\BibitemShut {NoStop}%
\bibitem [{\citenamefont {Kim}(1979)}]{kim_weak-interaction_1979}%
  \BibitemOpen
  \bibfield  {author} {\bibinfo {author} {\bibfnamefont {J.~E.}\ \bibnamefont
  {Kim}},\ }\href {https://doi.org/10.1103/PhysRevLett.43.103} {\bibfield
  {journal} {\bibinfo  {journal} {Physical Review Letters}\ }\textbf {\bibinfo
  {volume} {43}},\ \bibinfo {pages} {103} (\bibinfo {year} {1979})}\BibitemShut
  {NoStop}%
\bibitem [{\citenamefont {Dine}\ \emph {et~al.}(1981)\citenamefont {Dine},
  \citenamefont {Fischler},\ and\ \citenamefont
  {Srednicki}}]{dine_simple_1981}%
  \BibitemOpen
  \bibfield  {author} {\bibinfo {author} {\bibfnamefont {M.}~\bibnamefont
  {Dine}}, \bibinfo {author} {\bibfnamefont {W.}~\bibnamefont {Fischler}},\
  and\ \bibinfo {author} {\bibfnamefont {M.}~\bibnamefont {Srednicki}},\ }\href
  {https://doi.org/10.1016/0370-2693(81)90590-6} {\bibfield  {journal}
  {\bibinfo  {journal} {Physics Letters B}\ }\textbf {\bibinfo {volume}
  {104}},\ \bibinfo {pages} {199} (\bibinfo {year} {1981})}\BibitemShut
  {NoStop}%
\bibitem [{\citenamefont {Sikivie}(1983)}]{Sikivie_experimental_1983}%
  \BibitemOpen
  \bibfield  {author} {\bibinfo {author} {\bibfnamefont {P.}~\bibnamefont
  {Sikivie}},\ }\href {https://doi.org/10.1103/PhysRevLett.51.1415} {\bibfield
  {journal} {\bibinfo  {journal} {Phys. Rev. Lett.}\ }\textbf {\bibinfo
  {volume} {51}},\ \bibinfo {pages} {1415} (\bibinfo {year}
  {1983})}\BibitemShut {NoStop}%
\bibitem [{\citenamefont {{T. Braine, et al., (ADMX
  Collaboration)}}(2020)}]{admx_collaboration_extended_2020}%
  \BibitemOpen
  \bibfield  {author} {\bibinfo {author} {\bibnamefont {{T. Braine, et al.,
  (ADMX Collaboration)}}},\ }\href@noop {} {\bibfield  {journal} {\bibinfo
  {journal} {Physical Review Letters}\ }\textbf {\bibinfo {volume} {124}},\
  \bibinfo {pages} {101303} (\bibinfo {year} {2020})}\BibitemShut {NoStop}%
\bibitem [{\citenamefont {Eggemeier}\ \emph {et~al.}(2022)\citenamefont
  {Eggemeier}, \citenamefont {O'Hare}, \citenamefont {Pierobon}, \citenamefont
  {Redondo},\ and\ \citenamefont {Wong}}]{eggemeier_minivoids_2022}%
  \BibitemOpen
  \bibfield  {author} {\bibinfo {author} {\bibfnamefont {B.}~\bibnamefont
  {Eggemeier}}, \bibinfo {author} {\bibfnamefont {C.~A.~J.}\ \bibnamefont
  {O'Hare}}, \bibinfo {author} {\bibfnamefont {G.}~\bibnamefont {Pierobon}},
  \bibinfo {author} {\bibfnamefont {J.}~\bibnamefont {Redondo}},\ and\ \bibinfo
  {author} {\bibfnamefont {Y.~Y.~Y.}\ \bibnamefont {Wong}},\ }\href@noop {}
  {\bibinfo {title} {{Axion minivoids and implications for direct detection}}}
  (\bibinfo {year} {2022}),\ \Eprint {https://arxiv.org/abs/2212.00560}
  {arXiv:2212.00560 [hep-ph]} \BibitemShut {NoStop}%
\bibitem [{\citenamefont {Cameron}\ \emph {et~al.}(1993)\citenamefont
  {Cameron}, \citenamefont {Cantatore}, \citenamefont {Melissinos},
  \citenamefont {Ruoso}, \citenamefont {Semertzidis}, \citenamefont {Halama},
  \citenamefont {Lazarus}, \citenamefont {Prodell}, \citenamefont {Nezrick},
  \citenamefont {Rizzo},\ and\ \citenamefont {Zavattini}}]{Cameron_optic_1993}%
  \BibitemOpen
  \bibfield  {author} {\bibinfo {author} {\bibfnamefont {R.}~\bibnamefont
  {Cameron}}, \bibinfo {author} {\bibfnamefont {G.}~\bibnamefont {Cantatore}},
  \bibinfo {author} {\bibfnamefont {A.~C.}\ \bibnamefont {Melissinos}},
  \bibinfo {author} {\bibfnamefont {G.}~\bibnamefont {Ruoso}}, \bibinfo
  {author} {\bibfnamefont {Y.}~\bibnamefont {Semertzidis}}, \bibinfo {author}
  {\bibfnamefont {H.~J.}\ \bibnamefont {Halama}}, \bibinfo {author}
  {\bibfnamefont {D.~M.}\ \bibnamefont {Lazarus}}, \bibinfo {author}
  {\bibfnamefont {A.~G.}\ \bibnamefont {Prodell}}, \bibinfo {author}
  {\bibfnamefont {F.}~\bibnamefont {Nezrick}}, \bibinfo {author} {\bibfnamefont
  {C.}~\bibnamefont {Rizzo}},\ and\ \bibinfo {author} {\bibfnamefont
  {E.}~\bibnamefont {Zavattini}},\ }\href
  {https://doi.org/10.1103/PhysRevD.47.3707} {\bibfield  {journal} {\bibinfo
  {journal} {Phys. Rev. D}\ }\textbf {\bibinfo {volume} {47}},\ \bibinfo
  {pages} {3707} (\bibinfo {year} {1993})}\BibitemShut {NoStop}%
\bibitem [{\citenamefont {Zyla}\ \emph {et~al.}(2020)\citenamefont {Zyla} \emph
  {et~al.}}]{zyla_particle_2020}%
  \BibitemOpen
  \bibfield  {author} {\bibinfo {author} {\bibfnamefont {P.~A.}\ \bibnamefont
  {Zyla}} \emph {et~al.} (\bibinfo {collaboration} {Particle Data Group}),\
  }\bibfield  {journal} {\bibinfo  {journal} {Progress of Theoretical and
  Experimental Physics}\ }\textbf {\bibinfo {volume} {2020}},\ \href
  {https://doi.org/10.1093/ptep/ptaa104} {10.1093/ptep/ptaa104} (\bibinfo
  {year} {2020}),\ \bibinfo {note} {083C01}\BibitemShut {NoStop}%
\bibitem [{\citenamefont {Tam}\ and\ \citenamefont
  {Yang}(2012)}]{tam_production_2012}%
  \BibitemOpen
  \bibfield  {author} {\bibinfo {author} {\bibfnamefont {H.}~\bibnamefont
  {Tam}}\ and\ \bibinfo {author} {\bibfnamefont {Q.}~\bibnamefont {Yang}},\
  }\href {https://doi.org/10.1016/j.physletb.2012.08.050} {\bibfield  {journal}
  {\bibinfo  {journal} {Physics Letters B}\ }\textbf {\bibinfo {volume}
  {716}},\ \bibinfo {pages} {435} (\bibinfo {year} {2012})}\BibitemShut
  {NoStop}%
\bibitem [{\citenamefont {Raffelt}\ and\ \citenamefont
  {Stodolsky}(1988)}]{raffelt_mixing_1988}%
  \BibitemOpen
  \bibfield  {author} {\bibinfo {author} {\bibfnamefont {G.}~\bibnamefont
  {Raffelt}}\ and\ \bibinfo {author} {\bibfnamefont {L.}~\bibnamefont
  {Stodolsky}},\ }\href {https://doi.org/10.1103/PhysRevD.37.1237} {\bibfield
  {journal} {\bibinfo  {journal} {Physical Review D}\ }\textbf {\bibinfo
  {volume} {37}},\ \bibinfo {pages} {1237} (\bibinfo {year}
  {1988})}\BibitemShut {NoStop}%
\bibitem [{\citenamefont {Cregan}\ \emph {et~al.}(1999)\citenamefont {Cregan},
  \citenamefont {Mangan}, \citenamefont {Knight}, \citenamefont {Birks},
  \citenamefont {Russell}, \citenamefont {Roberts},\ and\ \citenamefont
  {Allan}}]{cregan_single_1999}%
  \BibitemOpen
  \bibfield  {author} {\bibinfo {author} {\bibfnamefont {R.~F.}\ \bibnamefont
  {Cregan}}, \bibinfo {author} {\bibfnamefont {B.~J.}\ \bibnamefont {Mangan}},
  \bibinfo {author} {\bibfnamefont {J.~C.}\ \bibnamefont {Knight}}, \bibinfo
  {author} {\bibfnamefont {T.~A.}\ \bibnamefont {Birks}}, \bibinfo {author}
  {\bibfnamefont {P.~S.~J.}\ \bibnamefont {Russell}}, \bibinfo {author}
  {\bibfnamefont {P.~J.}\ \bibnamefont {Roberts}},\ and\ \bibinfo {author}
  {\bibfnamefont {D.~C.}\ \bibnamefont {Allan}},\ }\href
  {https://doi.org/10.1126/science.285.5433.1537} {\bibfield  {journal}
  {\bibinfo  {journal} {Science}\ }\textbf {\bibinfo {volume} {285}},\ \bibinfo
  {pages} {1537} (\bibinfo {year} {1999})}\BibitemShut {NoStop}%
\bibitem [{\citenamefont {Russell}(2003)}]{russel_photonic_2003}%
  \BibitemOpen
  \bibfield  {author} {\bibinfo {author} {\bibfnamefont {P.}~\bibnamefont
  {Russell}},\ }\href {https://doi.org/10.1126/science.1079280} {\bibfield
  {journal} {\bibinfo  {journal} {Science}\ }\textbf {\bibinfo {volume}
  {299}},\ \bibinfo {pages} {358} (\bibinfo {year} {2003})}\BibitemShut
  {NoStop}%
\bibitem [{\citenamefont {Nikodem}(2020)}]{nikodem_laser_2020}%
  \BibitemOpen
  \bibfield  {author} {\bibinfo {author} {\bibfnamefont {M.}~\bibnamefont
  {Nikodem}},\ }\bibfield  {journal} {\bibinfo  {journal} {Materials}\ }\textbf
  {\bibinfo {volume} {13}},\ \href {https://doi.org/10.3390/ma13183983}
  {10.3390/ma13183983} (\bibinfo {year} {2020})\BibitemShut {NoStop}%
\bibitem [{\citenamefont {Michaille}\ \emph {et~al.}(2004)\citenamefont
  {Michaille}, \citenamefont {Taylor}, \citenamefont {Bennett}, \citenamefont
  {Shepherd}, \citenamefont {Jacobsen},\ and\ \citenamefont
  {Hansen}}]{Laurent_2004}%
  \BibitemOpen
  \bibfield  {author} {\bibinfo {author} {\bibfnamefont {L.~F.}\ \bibnamefont
  {Michaille}}, \bibinfo {author} {\bibfnamefont {D.~M.}\ \bibnamefont
  {Taylor}}, \bibinfo {author} {\bibfnamefont {C.~R.~H.}\ \bibnamefont
  {Bennett}}, \bibinfo {author} {\bibfnamefont {T.~J.}\ \bibnamefont
  {Shepherd}}, \bibinfo {author} {\bibfnamefont {C.}~\bibnamefont {Jacobsen}},\
  and\ \bibinfo {author} {\bibfnamefont {T.~P.}\ \bibnamefont {Hansen}},\ }in\
  \href {https://doi.org/10.1117/12.583481} {\emph {\bibinfo {booktitle}
  {Integrated Optical Devices, Nanostructures, and Displays}}},\ Vol.\ \bibinfo
  {volume} {5618},\ \bibinfo {editor} {edited by\ \bibinfo {editor}
  {\bibfnamefont {K.~L.}\ \bibnamefont {Lewis}}},\ \bibinfo {organization}
  {International Society for Optics and Photonics}\ (\bibinfo  {publisher}
  {SPIE},\ \bibinfo {year} {2004})\ pp.\ \bibinfo {pages} {30 --
  38}\BibitemShut {NoStop}%
\bibitem [{\citenamefont {Cao}\ \emph {et~al.}(2014)\citenamefont {Cao},
  \citenamefont {Jin}, \citenamefont {Yang},\ and\ \citenamefont
  {Ho}}]{cao_fiber_2014}%
  \BibitemOpen
  \bibfield  {author} {\bibinfo {author} {\bibfnamefont {Y.}~\bibnamefont
  {Cao}}, \bibinfo {author} {\bibfnamefont {W.}~\bibnamefont {Jin}}, \bibinfo
  {author} {\bibfnamefont {F.}~\bibnamefont {Yang}},\ and\ \bibinfo {author}
  {\bibfnamefont {H.~L.}\ \bibnamefont {Ho}},\ }\href
  {https://doi.org/10.1364/OE.22.013190} {\bibfield  {journal} {\bibinfo
  {journal} {Opt. Express}\ }\textbf {\bibinfo {volume} {22}},\ \bibinfo
  {pages} {13190} (\bibinfo {year} {2014})}\BibitemShut {NoStop}%
\bibitem [{\citenamefont {Triches}\ \emph {et~al.}(2015)\citenamefont
  {Triches}, \citenamefont {Brusch},\ and\ \citenamefont
  {Hald}}]{triches_portable_2015}%
  \BibitemOpen
  \bibfield  {author} {\bibinfo {author} {\bibfnamefont {M.}~\bibnamefont
  {Triches}}, \bibinfo {author} {\bibfnamefont {A.}~\bibnamefont {Brusch}},\
  and\ \bibinfo {author} {\bibfnamefont {J.}~\bibnamefont {Hald}},\ }\href
  {https://doi.org/10.1007/s00340-015-6224-8} {\bibfield  {journal} {\bibinfo
  {journal} {Applied Physics B}\ }\textbf {\bibinfo {volume} {121}},\ \bibinfo
  {pages} {251} (\bibinfo {year} {2015})}\BibitemShut {NoStop}%
\bibitem [{\citenamefont {Fini}\ \emph {et~al.}(2014)\citenamefont {Fini},
  \citenamefont {Nicholson}, \citenamefont {Mangan}, \citenamefont {Meng},
  \citenamefont {Windeler}, \citenamefont {Monberg}, \citenamefont {DeSantolo},
  \citenamefont {DiMarcello},\ and\ \citenamefont
  {Mukasa}}]{fini_polarization_2014}%
  \BibitemOpen
  \bibfield  {author} {\bibinfo {author} {\bibfnamefont {J.~M.}\ \bibnamefont
  {Fini}}, \bibinfo {author} {\bibfnamefont {J.~W.}\ \bibnamefont {Nicholson}},
  \bibinfo {author} {\bibfnamefont {B.}~\bibnamefont {Mangan}}, \bibinfo
  {author} {\bibfnamefont {L.}~\bibnamefont {Meng}}, \bibinfo {author}
  {\bibfnamefont {R.~S.}\ \bibnamefont {Windeler}}, \bibinfo {author}
  {\bibfnamefont {E.~M.}\ \bibnamefont {Monberg}}, \bibinfo {author}
  {\bibfnamefont {A.}~\bibnamefont {DeSantolo}}, \bibinfo {author}
  {\bibfnamefont {F.~V.}\ \bibnamefont {DiMarcello}},\ and\ \bibinfo {author}
  {\bibfnamefont {K.}~\bibnamefont {Mukasa}},\ }\href
  {https://doi.org/10.1038/ncomms6085} {\bibfield  {journal} {\bibinfo
  {journal} {Nature Communications}\ }\textbf {\bibinfo {volume} {5}},\
  \bibinfo {pages} {5085} (\bibinfo {year} {2014})}\BibitemShut {NoStop}%
\bibitem [{\citenamefont {Chen}\ \emph {et~al.}(2016)\citenamefont {Chen},
  \citenamefont {Wang}, \citenamefont {Hu}, \citenamefont {Shu},\ and\
  \citenamefont {Liu}}]{chen_single_2016}%
  \BibitemOpen
  \bibfield  {author} {\bibinfo {author} {\bibfnamefont {K.}~\bibnamefont
  {Chen}}, \bibinfo {author} {\bibfnamefont {C.}~\bibnamefont {Wang}}, \bibinfo
  {author} {\bibfnamefont {H.}~\bibnamefont {Hu}}, \bibinfo {author}
  {\bibfnamefont {X.}~\bibnamefont {Shu}},\ and\ \bibinfo {author}
  {\bibfnamefont {C.}~\bibnamefont {Liu}},\ }\href
  {https://doi.org/10.1109/LPT.2016.2608830} {\bibfield  {journal} {\bibinfo
  {journal} {IEEE Photonics Technology Letters}\ }\textbf {\bibinfo {volume}
  {28}},\ \bibinfo {pages} {2617} (\bibinfo {year} {2016})}\BibitemShut
  {NoStop}%
\bibitem [{\citenamefont {Taranta}\ \emph {et~al.}(2020)\citenamefont
  {Taranta}, \citenamefont {Numkam~Fokoua}, \citenamefont {Abokhamis~Mousavi},
  \citenamefont {Hayes}, \citenamefont {Bradley}, \citenamefont {Jasion},\ and\
  \citenamefont {Poletti}}]{Taranta_polarization_purity_2020}%
  \BibitemOpen
  \bibfield  {author} {\bibinfo {author} {\bibfnamefont {A.}~\bibnamefont
  {Taranta}}, \bibinfo {author} {\bibfnamefont {E.}~\bibnamefont
  {Numkam~Fokoua}}, \bibinfo {author} {\bibfnamefont {S.}~\bibnamefont
  {Abokhamis~Mousavi}}, \bibinfo {author} {\bibfnamefont {J.~R.}\ \bibnamefont
  {Hayes}}, \bibinfo {author} {\bibfnamefont {T.~D.}\ \bibnamefont {Bradley}},
  \bibinfo {author} {\bibfnamefont {G.~T.}\ \bibnamefont {Jasion}},\ and\
  \bibinfo {author} {\bibfnamefont {F.}~\bibnamefont {Poletti}},\ }\href@noop
  {} {\bibfield  {journal} {\bibinfo  {journal} {Nat. Photonics}\ }\textbf
  {\bibinfo {volume} {14}},\ \bibinfo {pages} {504} (\bibinfo {year}
  {2020})}\BibitemShut {NoStop}%
\bibitem [{\citenamefont {Mirizzi}\ and\ \citenamefont
  {Montanino}(2009)}]{mirizzi_stochastic_2009}%
  \BibitemOpen
  \bibfield  {author} {\bibinfo {author} {\bibfnamefont {A.}~\bibnamefont
  {Mirizzi}}\ and\ \bibinfo {author} {\bibfnamefont {D.}~\bibnamefont
  {Montanino}},\ }\href {https://doi.org/10.1088/1475-7516/2009/12/004}
  {\bibfield  {journal} {\bibinfo  {journal} {JCAP}\ }\textbf {\bibinfo
  {volume} {12}},\ \bibinfo {pages} {004}},\ \Eprint
  {https://arxiv.org/abs/0911.0015} {arXiv:0911.0015 [astro-ph.HE]}
  \BibitemShut {NoStop}%
\bibitem [{\citenamefont {De~Angelis}\ \emph {et~al.}(2011)\citenamefont
  {De~Angelis}, \citenamefont {Galanti},\ and\ \citenamefont
  {Roncadelli}}]{deangelis_relevance_2011}%
  \BibitemOpen
  \bibfield  {author} {\bibinfo {author} {\bibfnamefont {A.}~\bibnamefont
  {De~Angelis}}, \bibinfo {author} {\bibfnamefont {G.}~\bibnamefont
  {Galanti}},\ and\ \bibinfo {author} {\bibfnamefont {M.}~\bibnamefont
  {Roncadelli}},\ }\href {https://doi.org/10.1103/PhysRevD.84.105030}
  {\bibfield  {journal} {\bibinfo  {journal} {Phys. Rev. D}\ }\textbf {\bibinfo
  {volume} {84}},\ \bibinfo {pages} {105030} (\bibinfo {year} {2011})},\
  \bibinfo {note} {[Erratum: Phys.Rev.D 87, 109903 (2013)]},\ \Eprint
  {https://arxiv.org/abs/1106.1132} {arXiv:1106.1132 [astro-ph.HE]}
  \BibitemShut {NoStop}%
\bibitem [{\citenamefont {{V. Anastassopoulos, et al., (CAST
  Collaboration)}}(2017)}]{CAST_new_2017}%
  \BibitemOpen
  \bibfield  {author} {\bibinfo {author} {\bibnamefont {{V. Anastassopoulos, et
  al., (CAST Collaboration)}}} (\bibinfo {collaboration} {CAST}),\ }\href
  {https://doi.org/10.1038/nphys4109} {\bibfield  {journal} {\bibinfo
  {journal} {Nature Phys.}\ }\textbf {\bibinfo {volume} {13}},\ \bibinfo
  {pages} {584} (\bibinfo {year} {2017})}\BibitemShut {NoStop}%
\bibitem [{\citenamefont {Sokolov}\ and\ \citenamefont
  {Ringwald}(2022)}]{sokolov_electromagnetic_2022}%
  \BibitemOpen
  \bibfield  {author} {\bibinfo {author} {\bibfnamefont {A.~V.}\ \bibnamefont
  {Sokolov}}\ and\ \bibinfo {author} {\bibfnamefont {A.}~\bibnamefont
  {Ringwald}},\ }\href {https://doi.org/10.48550/arXiv.2205.02605} {\bibinfo
  {title} {Electromagnetic {Couplings} of {Axions}}} (\bibinfo {year}
  {2022})\BibitemShut {NoStop}%
\bibitem [{\citenamefont {Ding}\ \emph {et~al.}(2020)\citenamefont {Ding},
  \citenamefont {Fokoua}, \citenamefont {Bradley}, \citenamefont {Poletti},
  \citenamefont {Richardson},\ and\ \citenamefont {Slavik}}]{Ding_fabry_2020}%
  \BibitemOpen
  \bibfield  {author} {\bibinfo {author} {\bibfnamefont {M.}~\bibnamefont
  {Ding}}, \bibinfo {author} {\bibfnamefont {E.~R.~N.}\ \bibnamefont {Fokoua}},
  \bibinfo {author} {\bibfnamefont {T.~D.}\ \bibnamefont {Bradley}}, \bibinfo
  {author} {\bibfnamefont {F.}~\bibnamefont {Poletti}}, \bibinfo {author}
  {\bibfnamefont {D.~J.}\ \bibnamefont {Richardson}},\ and\ \bibinfo {author}
  {\bibfnamefont {R.}~\bibnamefont {Slavik}},\ }in\ \href
  {https://doi.org/10.1364/CLEO_SI.2020.SF2P.2} {\emph {\bibinfo {booktitle}
  {Conference on Lasers and Electro-Optics}}}\ (\bibinfo  {publisher} {Optica
  Publishing Group},\ \bibinfo {year} {2020})\ p.\ \bibinfo {pages}
  {SF2P.2}\BibitemShut {NoStop}%
\bibitem [{\citenamefont {Marcatili}\ and\ \citenamefont
  {Schmeltzer}(1964)}]{marcatili_hollow_1964}%
  \BibitemOpen
  \bibfield  {author} {\bibinfo {author} {\bibfnamefont {E.~A.~J.}\
  \bibnamefont {Marcatili}}\ and\ \bibinfo {author} {\bibfnamefont {R.~A.}\
  \bibnamefont {Schmeltzer}},\ }\href
  {https://doi.org/https://doi.org/10.1002/j.1538-7305.1964.tb04108.x}
  {\bibfield  {journal} {\bibinfo  {journal} {Bell System Technical Journal}\
  }\textbf {\bibinfo {volume} {43}},\ \bibinfo {pages} {1783} (\bibinfo {year}
  {1964})}\BibitemShut {NoStop}%
\bibitem [{\citenamefont {Rosa}\ \emph {et~al.}(2021)\citenamefont {Rosa},
  \citenamefont {Melli},\ and\ \citenamefont {Vincetti}}]{Rosa_2021}%
  \BibitemOpen
  \bibfield  {author} {\bibinfo {author} {\bibfnamefont {L.}~\bibnamefont
  {Rosa}}, \bibinfo {author} {\bibfnamefont {F.}~\bibnamefont {Melli}},\ and\
  \bibinfo {author} {\bibfnamefont {L.}~\bibnamefont {Vincetti}},\ }\href
  {https://doi.org/10.3390/fib9100058} {\bibfield  {journal} {\bibinfo
  {journal} {Fibers}\ }\textbf {\bibinfo {volume} {9}},\ \bibinfo {pages} {58}
  (\bibinfo {year} {2021})}\BibitemShut {NoStop}%
\bibitem [{\citenamefont {{COMSOL AB}}(2021)}]{comsol_multiphysics}%
  \BibitemOpen
  \bibfield  {author} {\bibinfo {author} {\bibnamefont {{COMSOL AB}}},\
  }\href@noop {} {\bibinfo {title} {{COMSOL Multiphysics}}},\ \bibinfo
  {howpublished} {\url{https://www.comsol.com/}} (\bibinfo {year} {2021}),\
  \bibinfo {note} {cOMSOL Multiphysics® v. 6.1.}\BibitemShut {Stop}%
\bibitem [{\citenamefont {Uebel}\ \emph {et~al.}(2016)\citenamefont {Uebel},
  \citenamefont {G{\"u}nendi}, \citenamefont {Frosz}, \citenamefont {Ahmed},
  \citenamefont {Edavalath}, \citenamefont {M{\'e}nard},\ and\ \citenamefont
  {Russell}}]{uebel_broadband_2016}%
  \BibitemOpen
  \bibfield  {author} {\bibinfo {author} {\bibfnamefont {P.}~\bibnamefont
  {Uebel}}, \bibinfo {author} {\bibfnamefont {M.~C.}\ \bibnamefont
  {G{\"u}nendi}}, \bibinfo {author} {\bibfnamefont {M.~H.}\ \bibnamefont
  {Frosz}}, \bibinfo {author} {\bibfnamefont {G.}~\bibnamefont {Ahmed}},
  \bibinfo {author} {\bibfnamefont {N.~N.}\ \bibnamefont {Edavalath}}, \bibinfo
  {author} {\bibfnamefont {J.-M.}\ \bibnamefont {M{\'e}nard}},\ and\ \bibinfo
  {author} {\bibfnamefont {P.~S.~J.}\ \bibnamefont {Russell}},\ }\href@noop {}
  {\bibfield  {journal} {\bibinfo  {journal} {Optics letters}\ }\textbf
  {\bibinfo {volume} {41}},\ \bibinfo {pages} {1961} (\bibinfo {year}
  {2016})}\BibitemShut {NoStop}%
\bibitem [{\citenamefont {Finger}\ \emph {et~al.}(2014)\citenamefont {Finger},
  \citenamefont {Joly}, \citenamefont {Weiss},\ and\ \citenamefont
  {Russell}}]{finger_accuracy_2014}%
  \BibitemOpen
  \bibfield  {author} {\bibinfo {author} {\bibfnamefont {M.~A.}\ \bibnamefont
  {Finger}}, \bibinfo {author} {\bibfnamefont {N.~Y.}\ \bibnamefont {Joly}},
  \bibinfo {author} {\bibfnamefont {T.}~\bibnamefont {Weiss}},\ and\ \bibinfo
  {author} {\bibfnamefont {P.~S.}\ \bibnamefont {Russell}},\ }\href
  {https://doi.org/10.1364/OL.39.000821} {\bibfield  {journal} {\bibinfo
  {journal} {Opt. Lett.}\ }\textbf {\bibinfo {volume} {39}},\ \bibinfo {pages}
  {821} (\bibinfo {year} {2014})}\BibitemShut {NoStop}%
\bibitem [{\citenamefont {Mathar}(2007)}]{Mathar_2007}%
  \BibitemOpen
  \bibfield  {author} {\bibinfo {author} {\bibfnamefont {R.~J.}\ \bibnamefont
  {Mathar}},\ }\href {https://doi.org/10.1088/1464-4258/9/5/008} {\bibfield
  {journal} {\bibinfo  {journal} {Journal of Optics A: Pure and Applied
  Optics}\ }\textbf {\bibinfo {volume} {9}},\ \bibinfo {pages} {470} (\bibinfo
  {year} {2007})}\BibitemShut {NoStop}%
\bibitem [{\citenamefont {Masum}\ \emph {et~al.}(2019)\citenamefont {Masum},
  \citenamefont {Aminossadati}, \citenamefont {Kizil},\ and\ \citenamefont
  {Leonardi}}]{Masum_fiber_2019}%
  \BibitemOpen
  \bibfield  {author} {\bibinfo {author} {\bibfnamefont {B.~M.}\ \bibnamefont
  {Masum}}, \bibinfo {author} {\bibfnamefont {S.~M.}\ \bibnamefont
  {Aminossadati}}, \bibinfo {author} {\bibfnamefont {M.~S.}\ \bibnamefont
  {Kizil}},\ and\ \bibinfo {author} {\bibfnamefont {C.~R.}\ \bibnamefont
  {Leonardi}},\ }\href {https://doi.org/10.1364/AO.58.000963} {\bibfield
  {journal} {\bibinfo  {journal} {Appl. Opt.}\ }\textbf {\bibinfo {volume}
  {58}},\ \bibinfo {pages} {963} (\bibinfo {year} {2019})}\BibitemShut
  {NoStop}%
\bibitem [{\citenamefont {Dicaire}\ \emph {et~al.}(2010)\citenamefont
  {Dicaire}, \citenamefont {Beugnot},\ and\ \citenamefont
  {Th{\'e}venaz}}]{Dicaire_pressurization_2010}%
  \BibitemOpen
  \bibfield  {author} {\bibinfo {author} {\bibfnamefont {I.}~\bibnamefont
  {Dicaire}}, \bibinfo {author} {\bibfnamefont {J.-C.}\ \bibnamefont
  {Beugnot}},\ and\ \bibinfo {author} {\bibfnamefont {L.}~\bibnamefont
  {Th{\'e}venaz}},\ }in\ \href {https://doi.org/10.1117/12.866465} {\emph
  {\bibinfo {booktitle} {Fourth European Workshop on Optical Fibre Sensors}}},\
  Vol.\ \bibinfo {volume} {7653},\ \bibinfo {editor} {edited by\ \bibinfo
  {editor} {\bibfnamefont {J.~L.}\ \bibnamefont {Santos}}, \bibinfo {editor}
  {\bibfnamefont {B.}~\bibnamefont {Culshaw}}, \bibinfo {editor} {\bibfnamefont
  {J.~M.}\ \bibnamefont {L{\'o}pez-Higuera}},\ and\ \bibinfo {editor}
  {\bibfnamefont {W.~N.}\ \bibnamefont {MacPherson}}},\ \bibinfo {organization}
  {International Society for Optics and Photonics}\ (\bibinfo  {publisher}
  {SPIE},\ \bibinfo {year} {2010})\ p.\ \bibinfo {pages} {76530L}\BibitemShut
  {NoStop}%
\end{thebibliography}%

\end{document}